\documentclass[aip,amsmath,amssymb,reprint]{revtex4-1}
\usepackage{graphicx}
\usepackage{dcolumn}
\usepackage{bm}
\usepackage[utf8]{inputenc}
\usepackage[T1]{fontenc}
\usepackage{mathptmx}
\usepackage{bm}
\usepackage{xcolor}

\newcommand\diff{\,\mathrm d}

\draft 

\begin{document}

\title{Mapping electromagnetic fields structure in plasma using a spin polarized electron beam} 



\author{X.Y. An}
\affiliation{Key Laboratory for Laser Plasmas (MoE) and School of Physics and Astronomy, Shanghai Jiao Tong University, Shanghai, 200240, China}
\affiliation{Zhi Yuan College, Shanghai Jiao Tong University, Shanghai 200240, China}
\affiliation{Collaborative Innovation Center of IFSA (CICIFSA), Shanghai Jiao Tong University, Shanghai 200240, China}
\author{M. Chen}
\email[]{minchen@sjtu.edu.cn}
\affiliation{Key Laboratory for Laser Plasmas (MoE) and School of Physics and Astronomy, Shanghai Jiao Tong University, Shanghai, 200240, China}
\affiliation{Collaborative Innovation Center of IFSA (CICIFSA), Shanghai Jiao Tong University, Shanghai 200240, China}
\author{J.X. Li}
\affiliation{Key Laboratory for Nonequilibrium Synthesis and Modulation of Condensed Matter (MOE), School of Science, Xi’an Jiao Tong University, Xi’an 710049, China}
\author{S.M. Weng}
\affiliation{Key Laboratory for Laser Plasmas (MoE) and School of Physics and Astronomy, Shanghai Jiao Tong University, Shanghai, 200240, China}
\affiliation{Collaborative Innovation Center of IFSA (CICIFSA), Shanghai Jiao Tong University, Shanghai 200240, China}
\author{F. He}
\affiliation{Key Laboratory for Laser Plasmas (MoE) and School of Physics and Astronomy, Shanghai Jiao Tong University, Shanghai, 200240, China}
\affiliation{Collaborative Innovation Center of IFSA (CICIFSA), Shanghai Jiao Tong University, Shanghai 200240, China}
\author{Z.M. Sheng}
\email[]{zmsheng@sjtu.edu.cn}
\affiliation{Key Laboratory for Laser Plasmas (MoE) and School of Physics and Astronomy, Shanghai Jiao Tong University, Shanghai, 200240, China}
\affiliation{Collaborative Innovation Center of IFSA (CICIFSA), Shanghai Jiao Tong University, Shanghai 200240, China}
\affiliation{SUPA, Department of Physics, University of Strathclyde, Glasgow G4 0NG, UK}
\affiliation{Cockcroft Institute, Sci-Tech Daresbury, Warrington, WA4 4AD, UK}
\author{J. Zhang}
\affiliation{Key Laboratory for Laser Plasmas (MoE) and School of Physics and Astronomy, Shanghai Jiao Tong University, Shanghai, 200240, China}
\affiliation{Collaborative Innovation Center of IFSA (CICIFSA), Shanghai Jiao Tong University, Shanghai 200240, China}

\date{\today}

\begin{abstract}
We propose a scheme to mapping electromagnetic fields structure in plasma by using a spin polarized relativistic electron beam. Especially by using Particle-in-Cell (PIC) and electron spin tracing simulations, we have successfully reconstructed a plasma wakefield from the spin evolution of a transmitted electron beam. Electron trajectories of the probe beam are obtained from PIC simulations, and the spin evolutions during the beam propagating through the fields are calculated by a spin tracing code. The reconstructed fields illustrate the main characters of the original fields, which demonstrates the feasibility of fields detection by use of spin polarized relativistic electron beams.
\end{abstract}

\pacs{}

\maketitle 

\section{Introduction}
Nowadays, spin polarized electron beams with relativistic energy can be produced in several ways. In typical ways, electrons can be polarized via a spin filter\cite{spinfilter} or a beam splitter\cite{beamsplitter} and in a storage ring electrons can be polarized through radiative polarization (Sokolov-Ternov effect)\cite{produce1}. Alternatively, one can obtain polarized electrons from pre-polarized gas\cite{fromgas1} or photo-cathode\cite{cathode} and then accelerate them. It is also found that a polarized electron beam can be produced from nonlinear Compton scattering with an elliptically polarized laser\cite{EPlaser}. Along with the rapid development of ultrashort ultraintense laser technology and laser plasma wakefield acceleration scheme, tabletop compact spin polarized electron accelerators may be reality in the near future\cite{spinlwfa,bichromatic}.

Spin polarized electron beams with relativistic energy have now found wide applications. They can be used to study molecular and atomic structure\cite{molecularstruct}. Besides, polarized electron beams can also be employed to study high energy physics, such as probing nuclear structures\cite{nuclearstruct, nuclearstruct2}, observing parity nonconservation\cite{parity} and generating polarized photons\cite{polaphoton} and positrons\cite{polapositron1,polapositron2}. Furthermore, new physics beyond Standard Model may be discovered via polarized electron beams\cite{smmodel}. In this paper, we propose to use such kind of beam as a probe beam to diagnose plasma fields which are important in laser plasma studies such as inertial confinement fusion concept and plasma wakefield accelerator\cite{tajima,pisinchen}, but are also extremely difficult to be detected. To demonstrate its feasibility, in the current study we take plasma wakefields as the detected fields.

Plasma wakefield accelerators have been considered as next generation high energy  accelerator\cite{tajima,pisinchen,Esarey,protondrive}. GeV electron beams have been successfully accelerated within centimeter acceleration length from laser driven wakefields\cite{Gonsalves,Leemans,GeV3} and tens of GeV electron beams have been accelerated within meter scale acceleration length from beam driven wakefields\cite{42gev}. In order to get electron beam with high quality, it is important that the fast moving wakefields can be measured. Different detection approaches have been developed, such as frequency domain holography\cite{snapshots,FDH1,FDH2}, frequency-domain tomography\cite{frequencydiag1,frequencydiag2} or direct detection by a few-cycle probe pulse\cite{fewcyclediag}. Relativistic electron beam can also be used to detect the wakefields and other fields induced in plasma through the variation of the electron beam density distribution after transmission through the wake\cite{densitydiag,bunchdiag}.

In this paper, we extend the previous electron beam detection methods by using a spin polarized electron beam. As we will see that the field information will not only be projected into the density profile of the probe beam, but also imprint on its spin evolution. The latter can be used to reconstruct the fields.

\section{Physical Model of spin evolution and simulation setup }

\subsection{Physical model of spin dynamics}
We start by showing the physical models of spin evolution in electromagnetic fields. Usually the spin can change in two different ways. One is called the Sokolov-Ternov effect\cite{produce1}. In this way, the energy differs when the spin of the electron is parallel or antiparallel to an external magnetic field. The energy of the state in which the spin is antiparallel to the magnetic field is lower. Thus the electron may radiate a photon and transform the state from parallel to antiparallel. By such way it takes several minutes or longer to build up polarization in the storage ring\cite{STtime}. Therefore, in our femtosecond time scale case, Sokolov-Ternov effect can be neglected.

Besides Sokolov-Ternov effect, the spin can also precess in an electromagnetic field. In a non-relativistic case, it precesses in a quite simple form:
\begin{equation}
	\frac{\diff\bm s}{\diff t}=\bm\mu\times\bm B=-g\frac{\bm s\times\bm B}{2}
\end{equation}
where $\mu$ is electron intrinsic magnetic moment and $g\approx 2.00232$ is Landé g-factor of an electron\cite{gfactor1,gfactor2}.

However, in a relativistic case, the precession is more complicated and follows the so called T-BMT equation\cite{Thomas,Thomas2,TBMTequation,spinradiation}:
\begin{gather}
	\frac{\diff \bm s}{\diff t}=\bm\Omega\times\bm s=\left(\bm\Omega_T+\bm\Omega_a\right)\times\bm s \\
	\bm\Omega_T=\frac{e}{m}\left(\frac{1}{\gamma}\bm B-\frac{1}{\gamma+1}\frac{\bm v}{c^2}\times\bm E\right) \\
	\bm\Omega_a=a_e\frac{e}{m}\left[\bm B-\frac{\gamma}{\gamma+1}\frac{\bm v}{c^2}\left(\bm v\cdot\bm B\right)-\frac{\bm v}{c^2}\times\bm E\right]
\end{gather}
where $a_e=(g-2)/2=1.16\times10^{-3}$ and $\gamma, \bm v$ are the Lorentz factor and velocity of the electron, respectively.

In later discussion, the spin evolution of the probe beam will mainly precess as the T-BMT equation describes. Notice that T-BMT equation describes how the spin of a single electron precesses while in PIC simulation one macro-particle contains about $10^{10}$ electrons. Each electron in one macro-particle satisfies its own T-BMT equation:
\begin{equation}
	\frac{\diff \bm s_i}{\diff t}=\bm\Omega\left(\bm v_i,\bm E_i,\bm B_i\right)\times\bm s_i
	\label{eq:particleTBMT}
\end{equation}
where $\bm v_i$, $\bm E_i$, $\bm B_i$, $\bm s_i$ and $\bm\Omega\left(\bm v_i,\bm E_i,\bm B_i\right)$ are the velocity, electric field, magnetic field, spin and precession angular velocity of each electron in the macro-particle, respectively.

In PIC simulation, we assume all the electrons in the same macro-particle feel the same electromagnetic fields and make the same precession. Therefore, the equations of Eq.~(\ref{eq:particleTBMT}) can be combined as:
\begin{equation}
\begin{split}
	\frac{\diff \bm S}{\diff t}&=\frac{\diff (\sum \bm s_i)}{\diff t} \\
	&=\sum\left(\frac{\diff s_i}{\diff t}\right) \\
	&=\sum\left[\bm\Omega(\bm v,\bm E,\bm B)\times\bm s_i\right] \\
	&=\bm\Omega(\bm v,\bm E,\bm B)\times\sum \bm s_i \\
	&=\bm\Omega(\bm v,\bm E,\bm B)\times\bm S
\end{split}
\end{equation}
This means that T-BMT equation can also describe the spin precession of the macro-particle which represents the spin precession of the electrons in the macro-particle as well.

\begin{figure}[ht]
	\centering
	\includegraphics[scale=0.21]{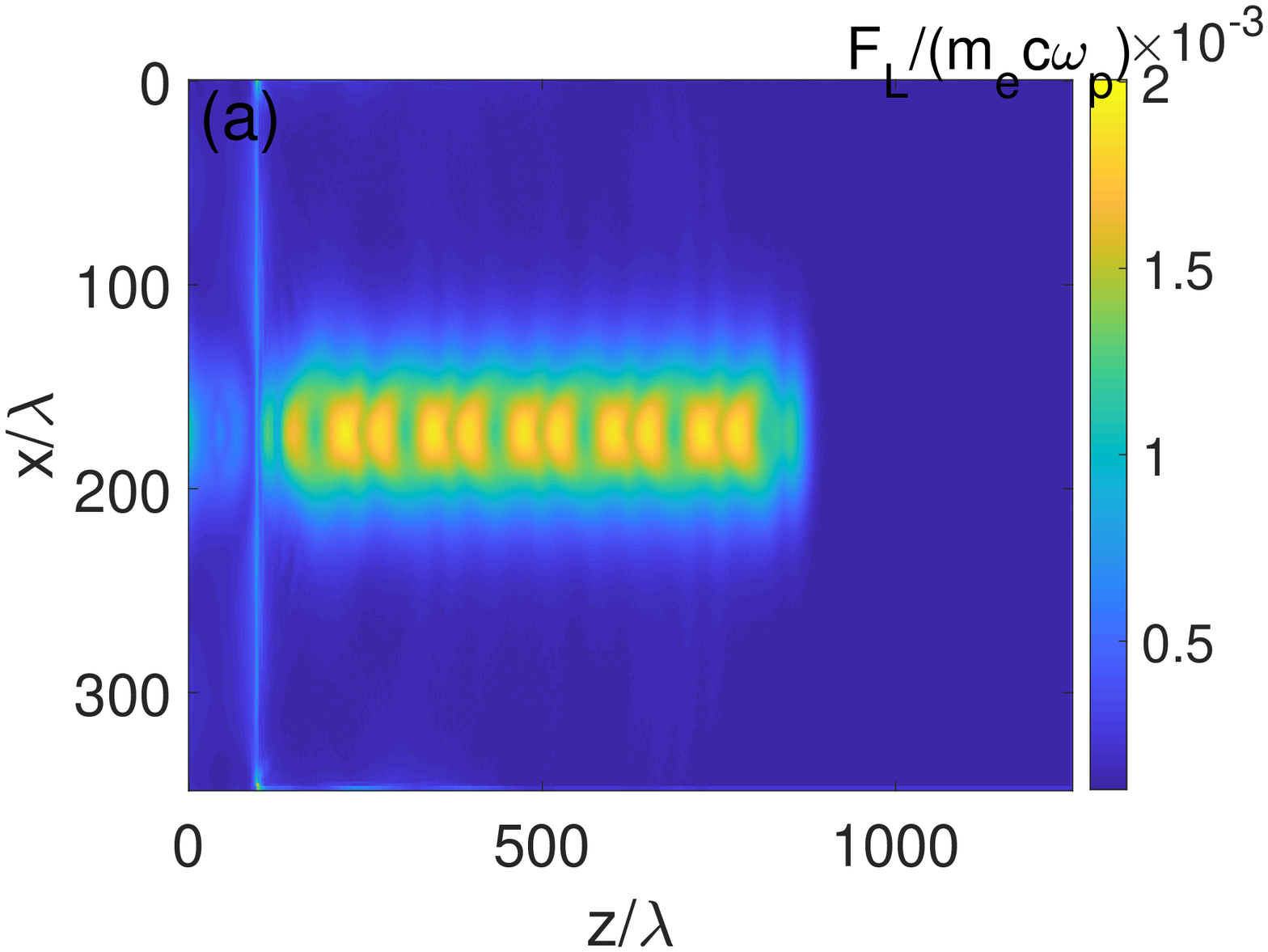}
	\includegraphics[scale=0.21]{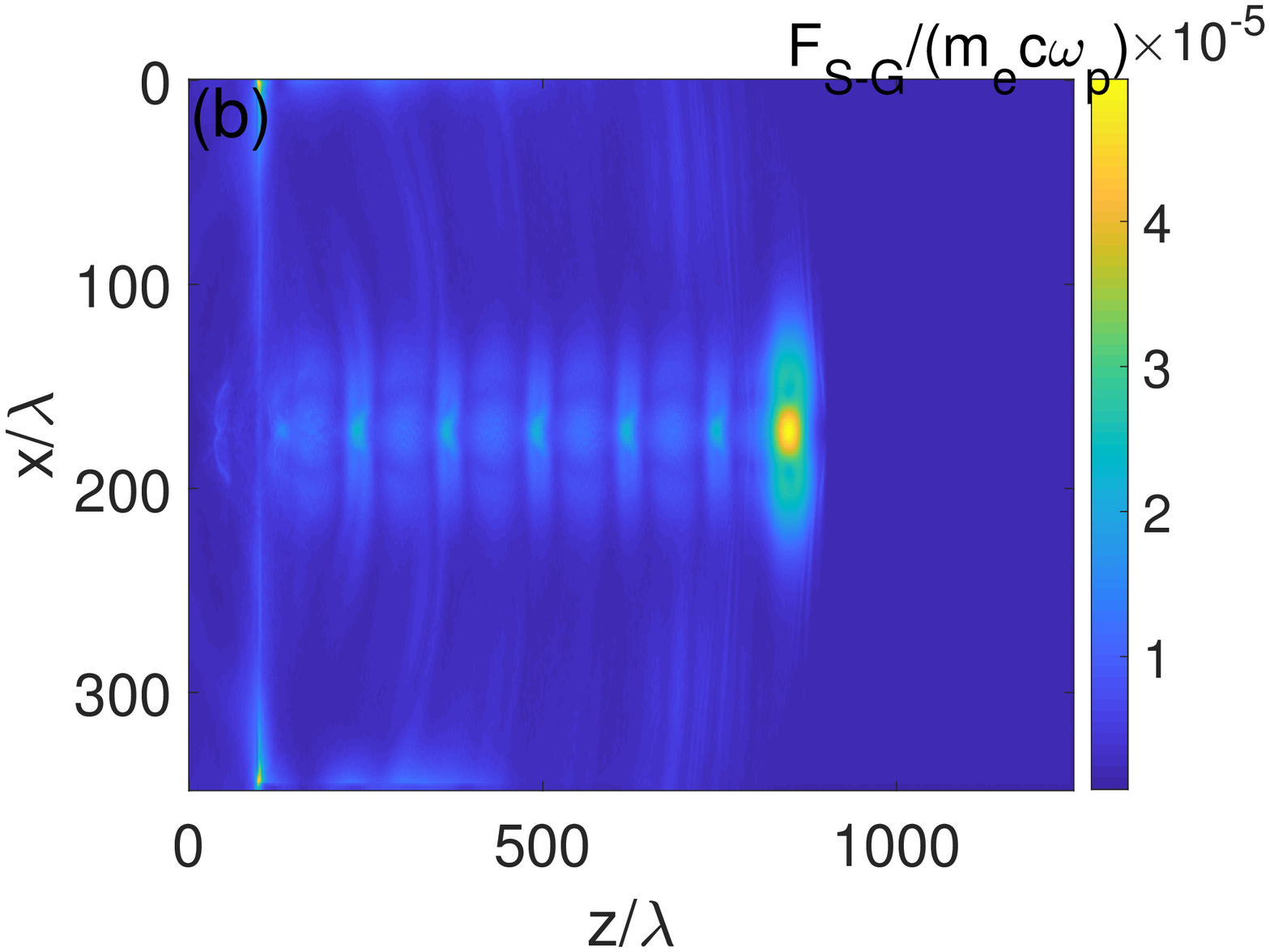}
	\caption{The Lorentz force (a) and the Stern-Gerlach force (b) felt by the electrons in the probe beam. It shows that the Lorentz force is about $10^{-3}$ and the Stern-Gerlach force is about $10^{-5}$. The detailed simulation parameters are introduced in the next section. Compared with the Lorentz force, the Stern-Gerlach force is negligible.}
	\label{fig:SGcheck}
\end{figure}

One the other hand, there is still a problem when we separate the movement and spin precession in different programs. We get the particle motion from the PIC simulation without considering the spin effect and calculate the spin precession from the spin evolution code with the fields tracing from the PIC simulation. Since the electrons are moving in an uneven magnetic field, they will feel not only the Lorentz force:
\begin{equation}
	F_{\rm L}=-e\left(\bm E+\bm v\times \bm B\right)
\end{equation}
but also the Stern-Gerlach force:
\begin{equation}
	F_{\rm{S-G}}=\nabla\left(\bm\mu\cdot\bm B\right)
\end{equation}
In this way, the spin of the electrons will affect the force they feel and thus change the motion and trajectory. However, in the relativistic case, the Stern-Gerlach force is negligible compared with the Lorentz force\cite{SGneglect,SGneglect2}. Also we checked the validity of this approximation in our simulations, the results are shown in Fig.~\ref{fig:SGcheck}.
\subsection{Numerical model of spin dynamics}
\begin{figure}[ht]
	\centering
	\includegraphics[scale=1.6]{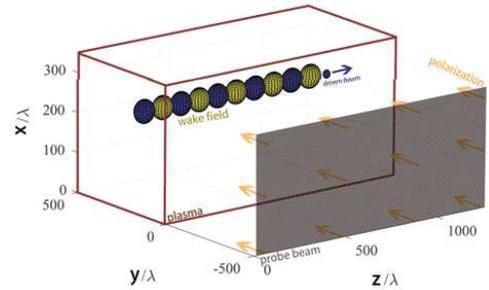}
	\caption{Schematic of the wakefield detection with a polarized electron beam probe and the numerical simulation configuration, where the probe is a slab beam propagating along the y-direction.}
	\label{fig:scheme}
\end{figure}

After introducing the physical model, we test our scheme through computational simulations. A relativistic electron beam is firstly used to drive a wakefield in underdense plasma. At the same time, a spin polarized relativistic electron probe beam with a cross section larger than the wakefield period is transversely incident to the wakefield from the outside of the plasma, as Fig.~\ref{fig:scheme} shows. After the probe beam transmits through the wakefield, the distribution of the electron spins will be modulated and such modulation is used to reconstruct the wakefield.

The wake excitation and probe beam transmission processes are simulated by three dimensional PIC simulation with the code OSIRIS\cite{osiris}. The simulation box is [0,$1000\,\mu\text{m}]\times[-400\,\mu\text{m},400\,\mu\text{m}]\times[0,278\,\mu\text{m}]$ and it is divided in grids as $1000\times 800\times 300$ in the longitudinal and two transverse directions, respectively. The plasma has a uniform density profile $n_p=1.1\times 10^{17}\,\text{cm}^{-3}$ in the region for $z>80\,\mu\text{m}$ and $y>0$. An electron beam with energy of 1\,GeV is used to drive the wakefield along the $z$ direction which will be detected by the probe electron beam. The driver beam has Gaussian density profiles both along the longitudinal and transverse directions with its peak density $n_b=0.05n_p$. The waist of the beam is $\sigma_z=20\,\mu\text{m},\sigma_r=17.8\,\mu\text{m}$. Such a driver beam can excite a linear wake. The probe beam is assumed to be a slab distributed in the area of $z\in[80\,\mu\text{m},400\,\mu\text{m}]$, $y\in[-398\,\mu\text{m},-400\,\mu\text{m}]$ and all the $x$ direction. The energy of the probe beam is 200\,MeV and there is no energy spread and emittance. The probe beam is initially completely polarized along $y$ direction. In the simulation study, the spatial and temporal units are normalized according to a laser wavelength of $\lambda=0.8 \mu$m and period of $T=2.67$ fs, respectively.

\begin{figure}[ht]
	\centering
	\includegraphics[scale=0.21]{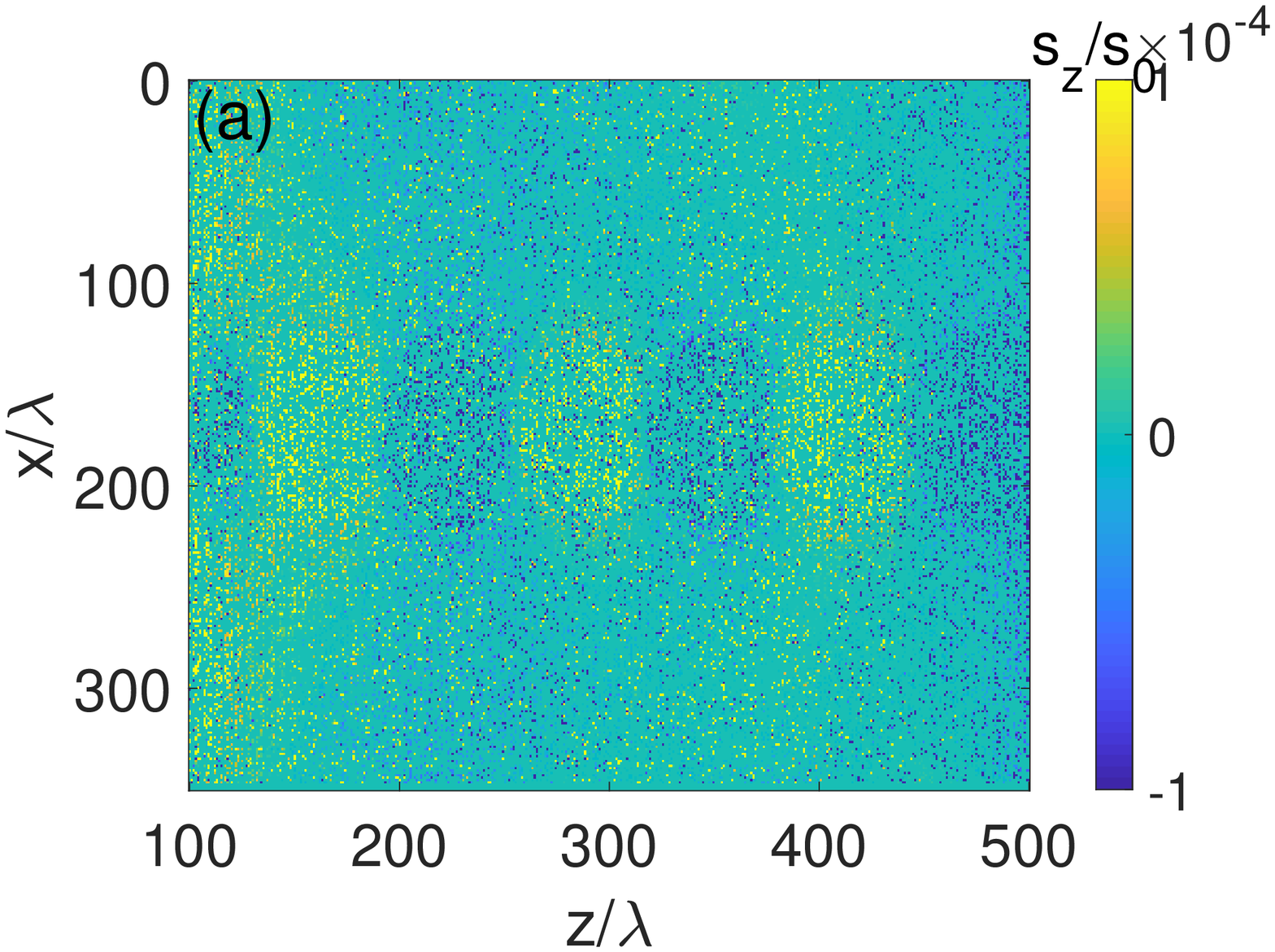}
	\includegraphics[scale=0.21]{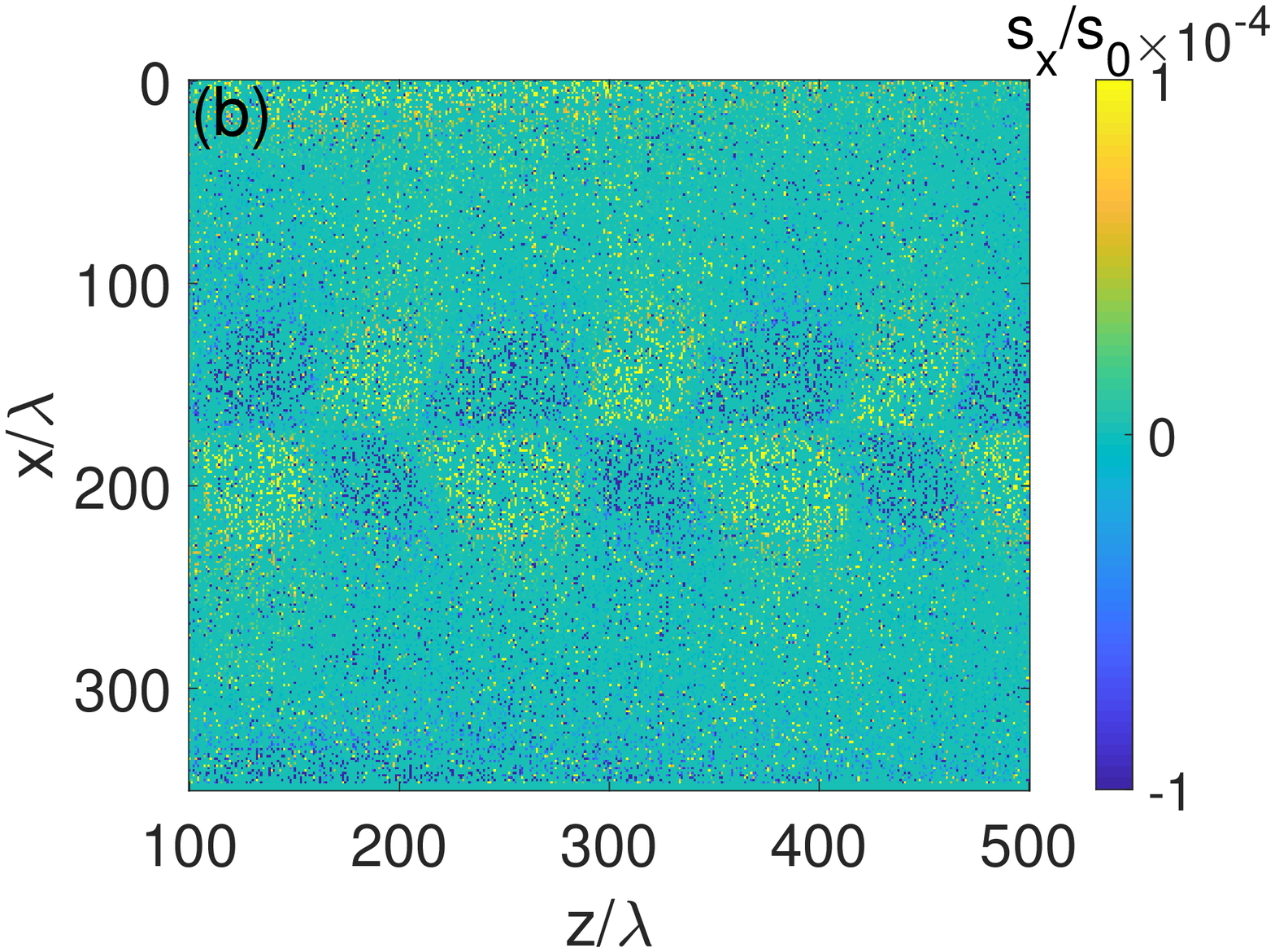}
	\caption{(a) The distribution of $s_z$ in the plane $x$-$z$ after the probe beam transmits through the wakefields.  (b) The distribution of $s_x$ in the plane $x$-$z$ after the probe beam transmits through the wakefields. Here the wake is propagating along the $z$ direction.}
	\label{fig:spindist}
\end{figure}

In the PIC simulations here, there is not yet a module added to consider the spin evolution. According to the above discussions, we can use the trajectory and fields data of each electron from PIC simulation and then calculate its spin evolution by a separate spin precession code. We have developed a parallel program to compute spin evolution for each electron in the probe beam along its trajectory by using the fields information traced from the PIC simulations. After the whole probe beam traveling through the wakefields, the spin distribution of the beam is calculated and a typical result is shown in Fig.~\ref{fig:spindist}. As one can see that a wakefield like distribution appears.

\section{Wakefield reconstruction}
\subsection{Reconstruction of field structures in the longitudinal direction}
\begin{figure}[ht]
	\centering
	\includegraphics[scale=0.21]{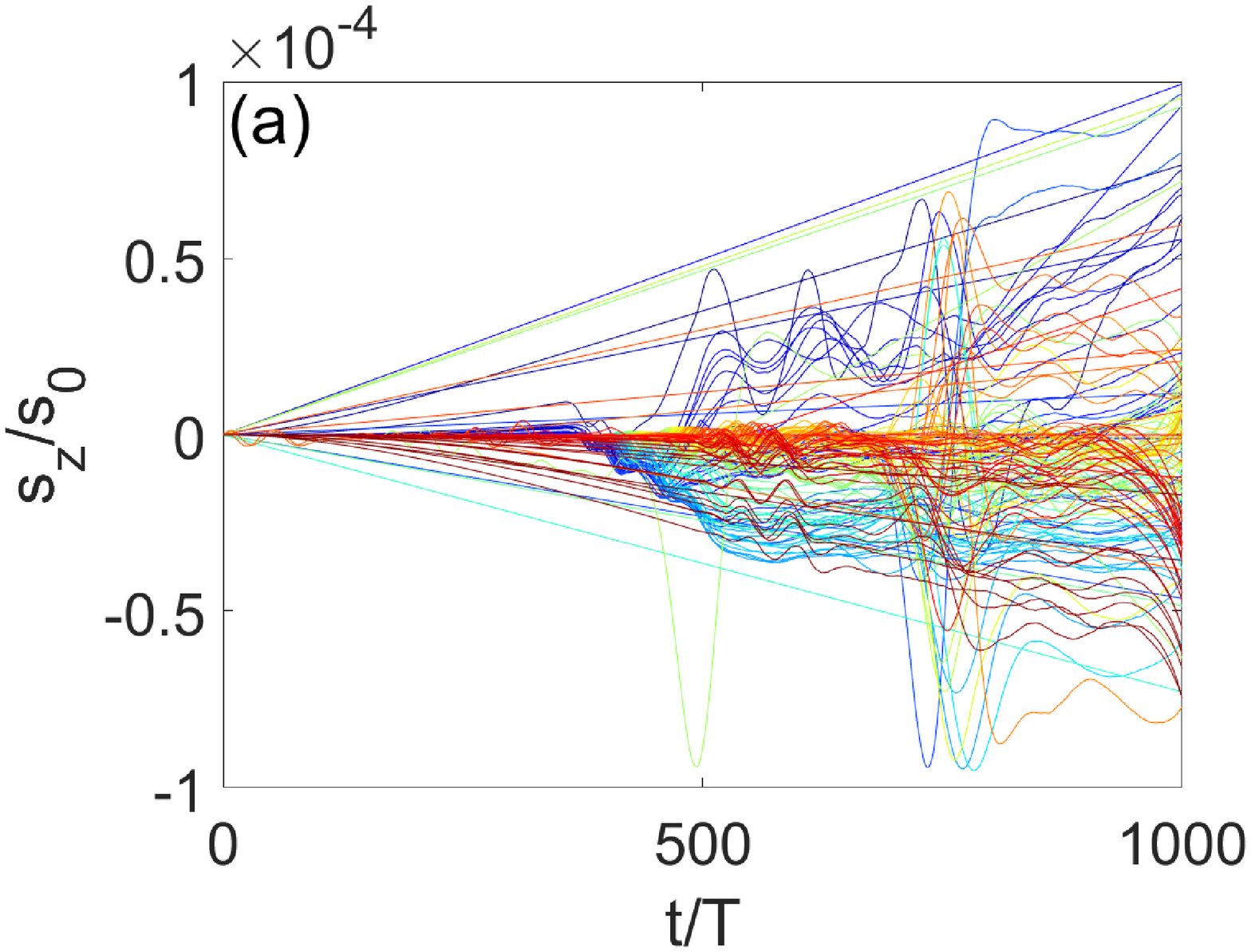}
	\includegraphics[scale=0.21]{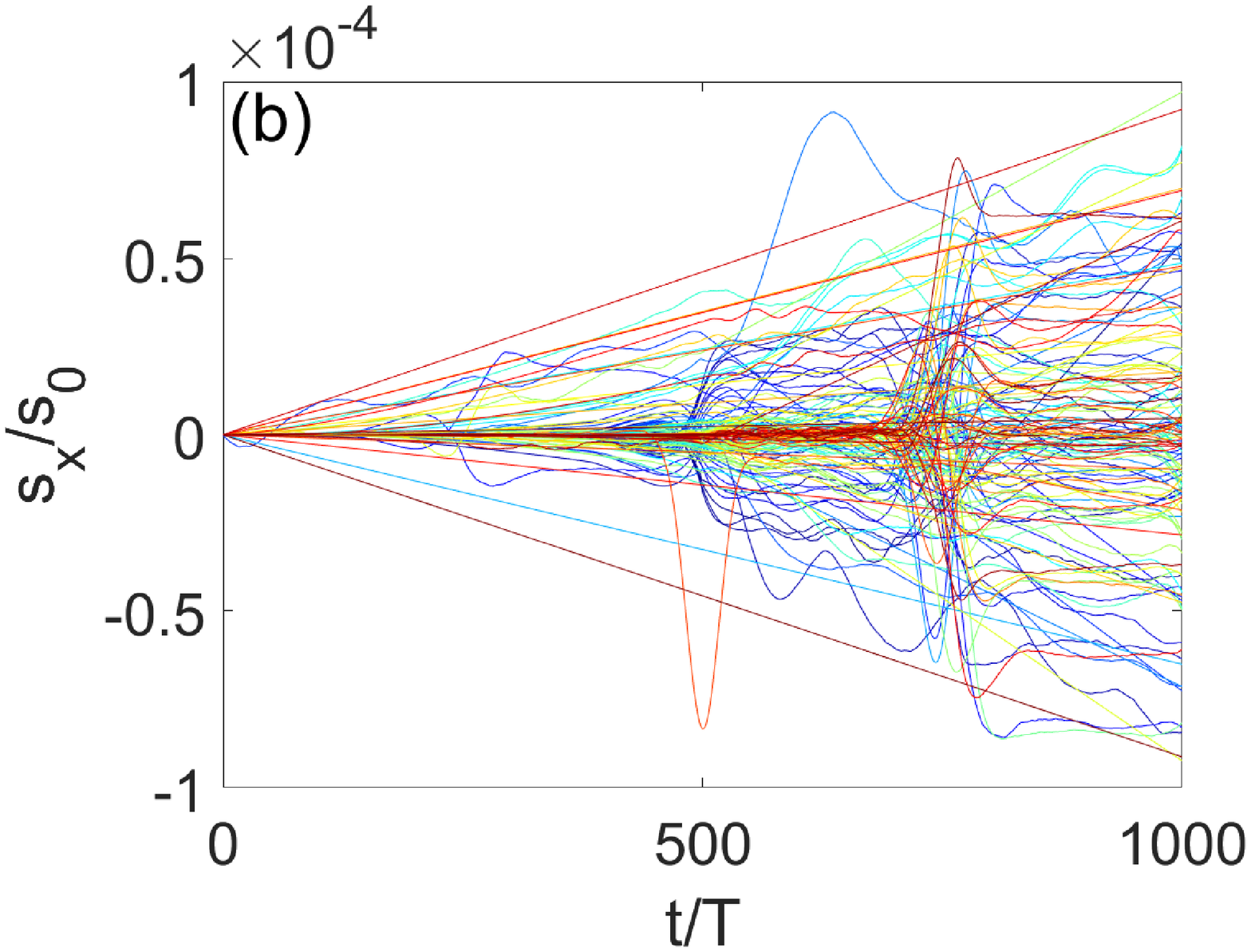}
	\caption{The temporal evolutions of $s_z$ (a) and $s_x$ (b). Each color represents different electron.\label{fig:timedependent}}
\end{figure}

Before reconstructing the wakefield from the probe beam's spin evolution, we firstly study the case of a single electron. Fig.~\ref{fig:timedependent} shows the temporal evolution of the single electron's spin precession. We have randomly selected one percent of the whole electrons to show their spin evolution.

\begin{figure}[ht]
	\centering
	\includegraphics[scale=0.21]{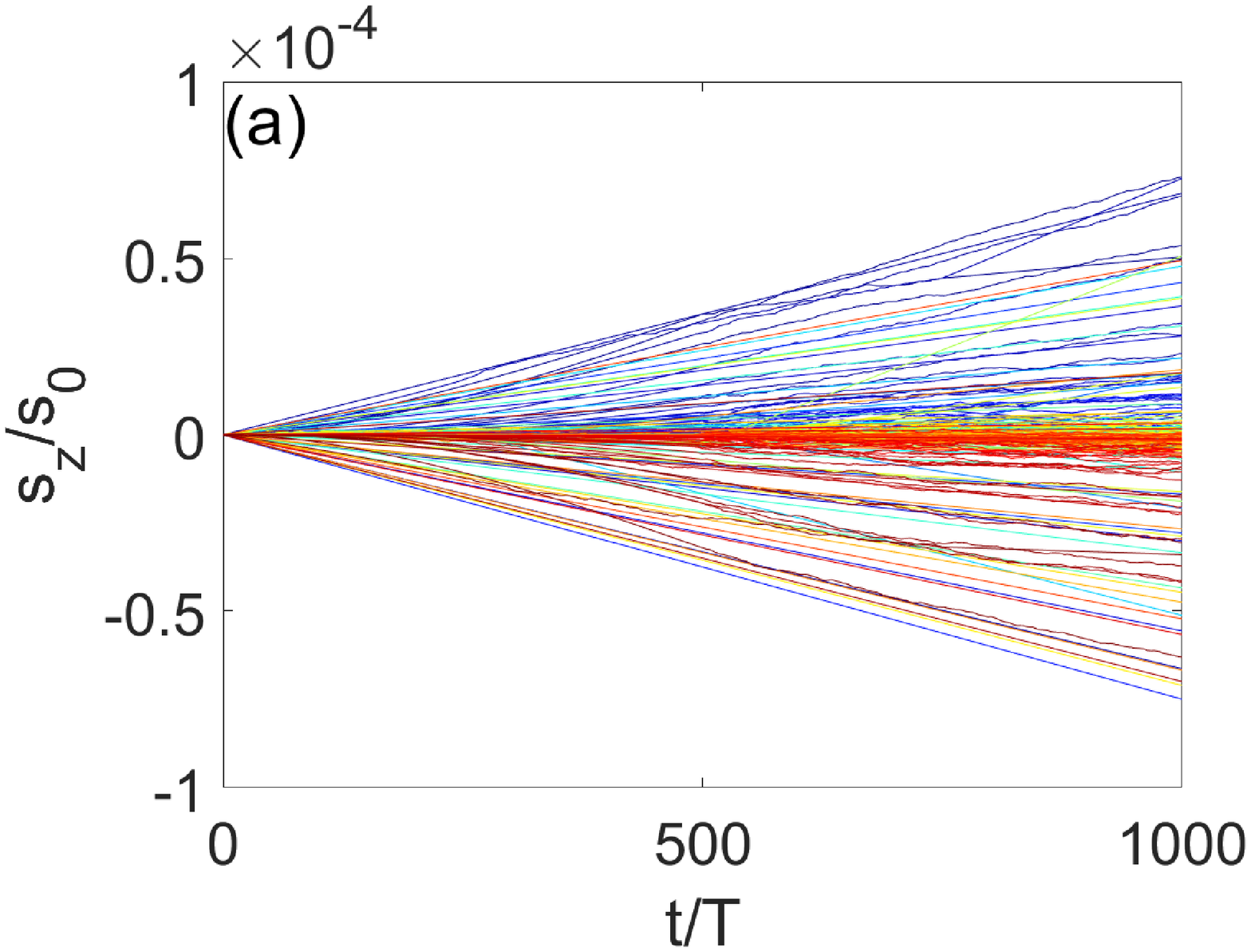}
	\includegraphics[scale=0.21]{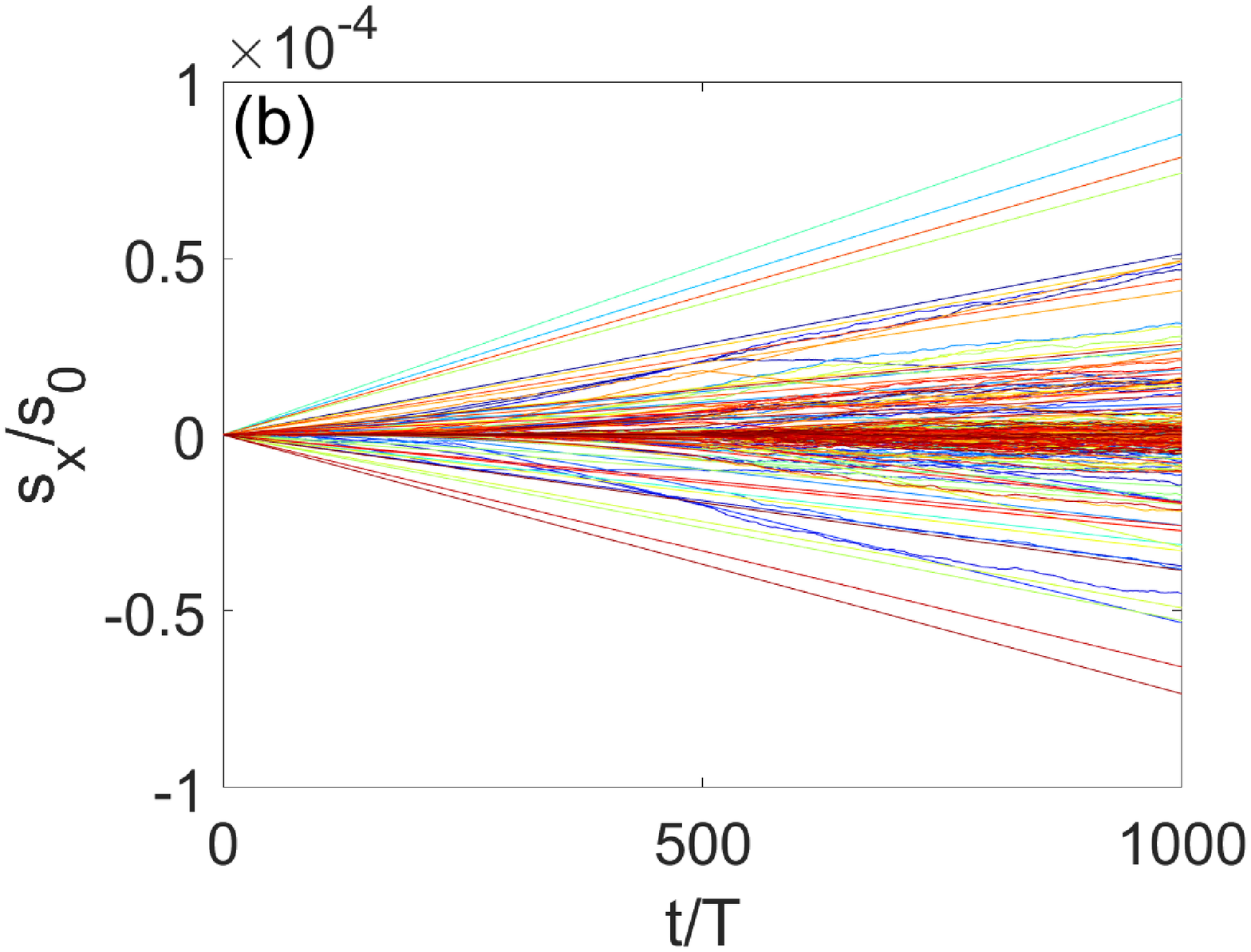}
	\caption{The $s_z$ (a) and $s_x$ (b) evolutions of the probe beam in the PBFP simulation. All the electrons spin evolution in this case show linear character. \label{fig:TDcontrol}}
\end{figure}

The probe beam touches the wakefield at about $t=700T$ and so it is reasonable to see that some of the electrons change their spin directions at that time. There are also some of electrons changing their spins at about $t=500T$. It is because that at this time the probe beam begins to enter into the plasma. The wakefield driven by the probe beam itself also induces the spin precession. Besides these two kinds of electrons, there are still some electrons whose spins vary almost linearly. This is because they are at the boundary of the probe beam and they interact with the self-generated electromagnetic fields by the probe beam. We have confirmed this boundary effects through a probe-beam-free-propagation (PBFP) simulation as shown in Fig.~\ref{fig:TDcontrol}. In this simulation, there is no driver beam exciting the plasma wake, while all the other plasma and probe beam parameters are as the same as before. In Fig.\ref{fig:TDcontrol}, we can see the spins of all the electrons evolve linearly. The more outer electrons, the more spin precession, which is consistent with the boundary effects.

\begin{figure}[ht]
	\centering
	\includegraphics[scale=0.21]{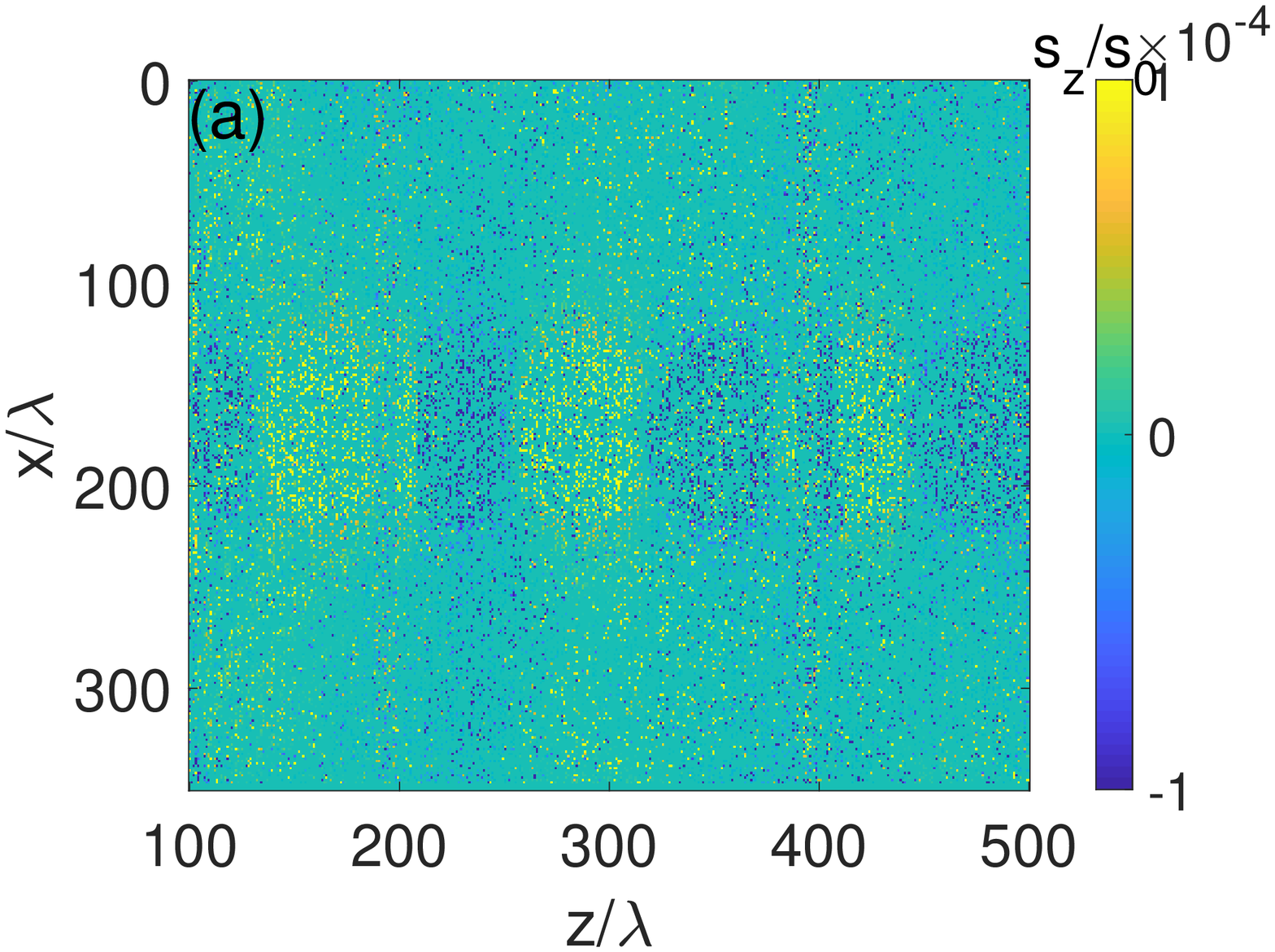}
	\includegraphics[scale=0.21]{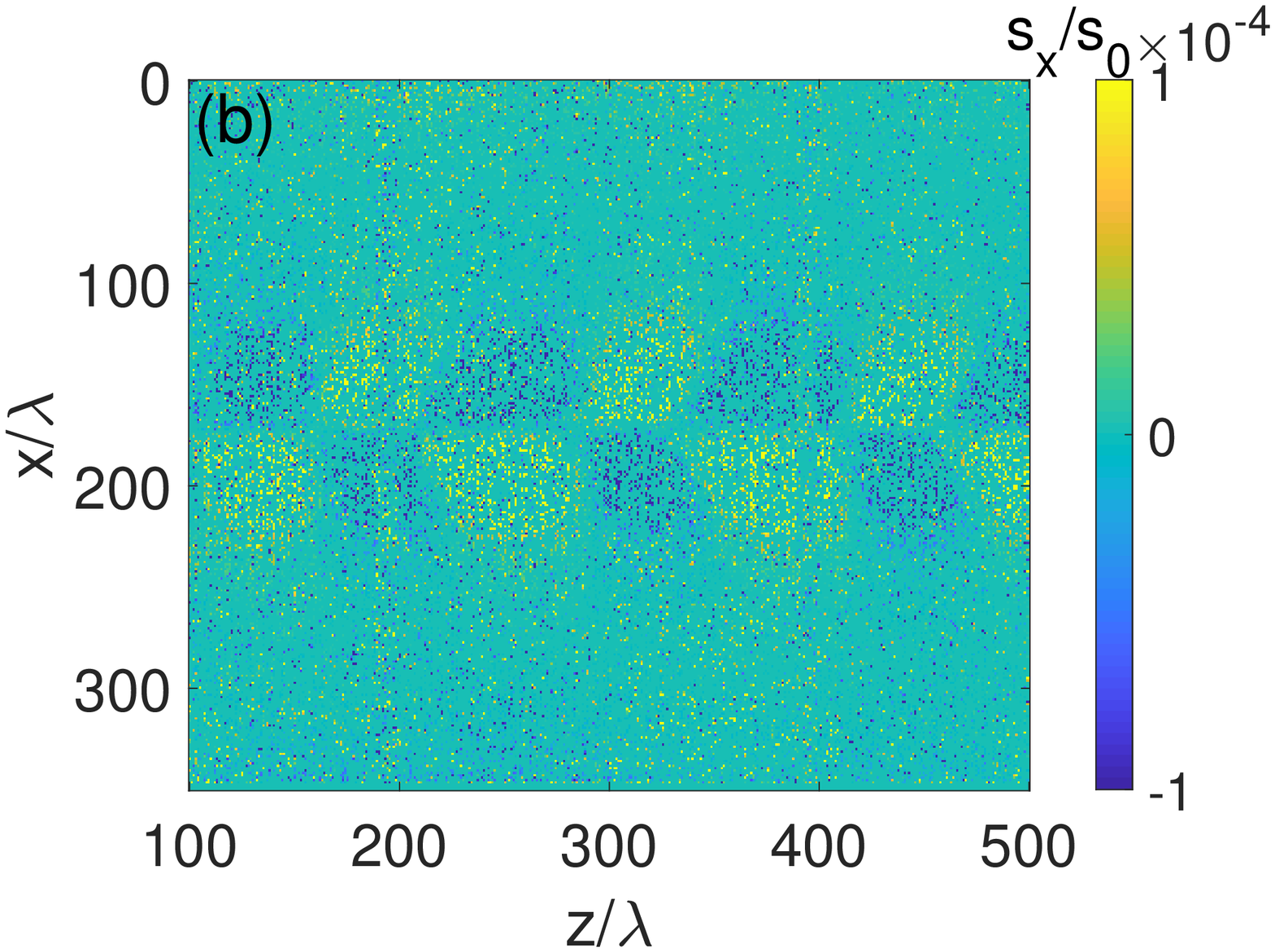}
	\includegraphics[scale=0.21]{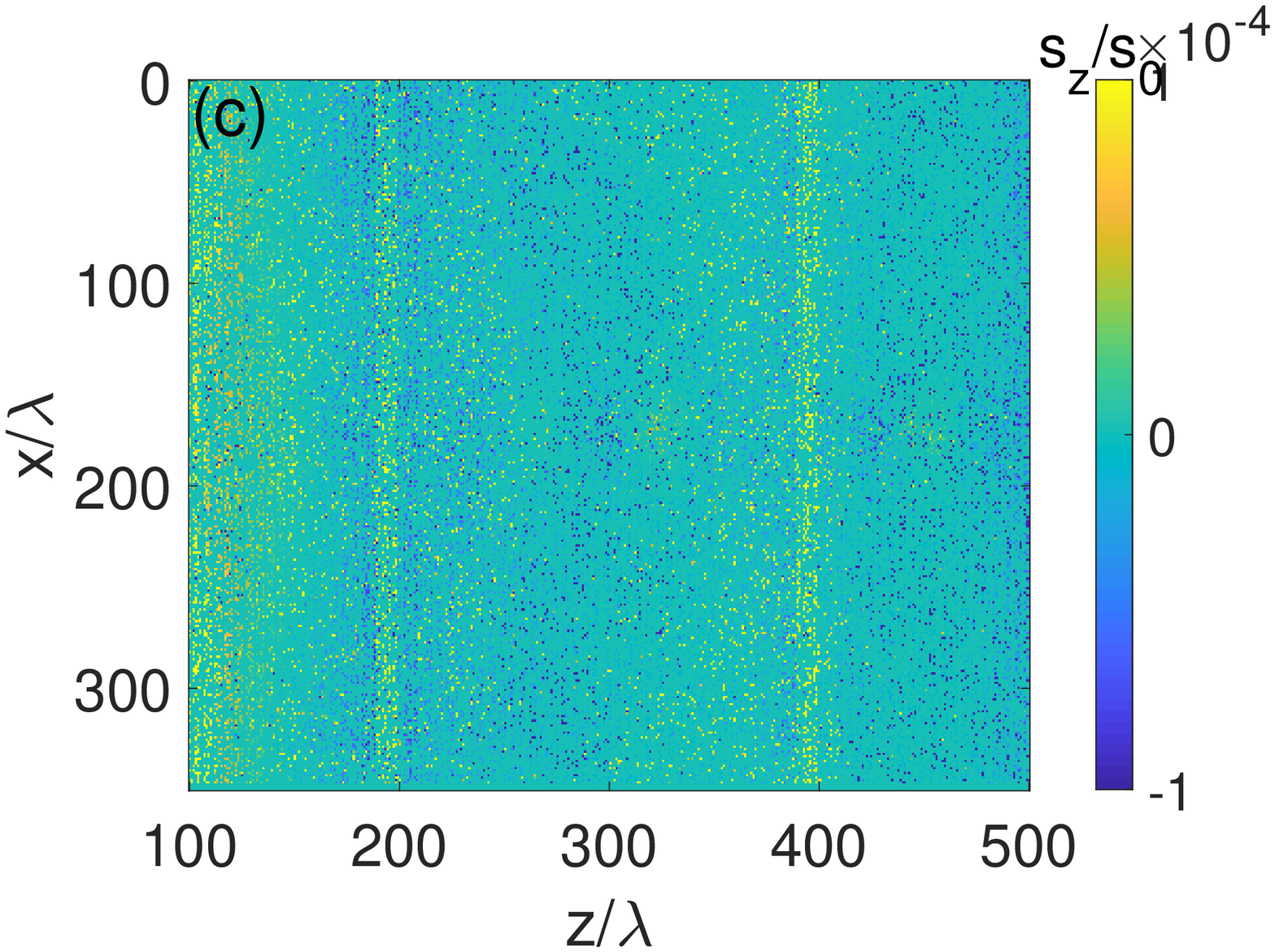}
	\includegraphics[scale=0.21]{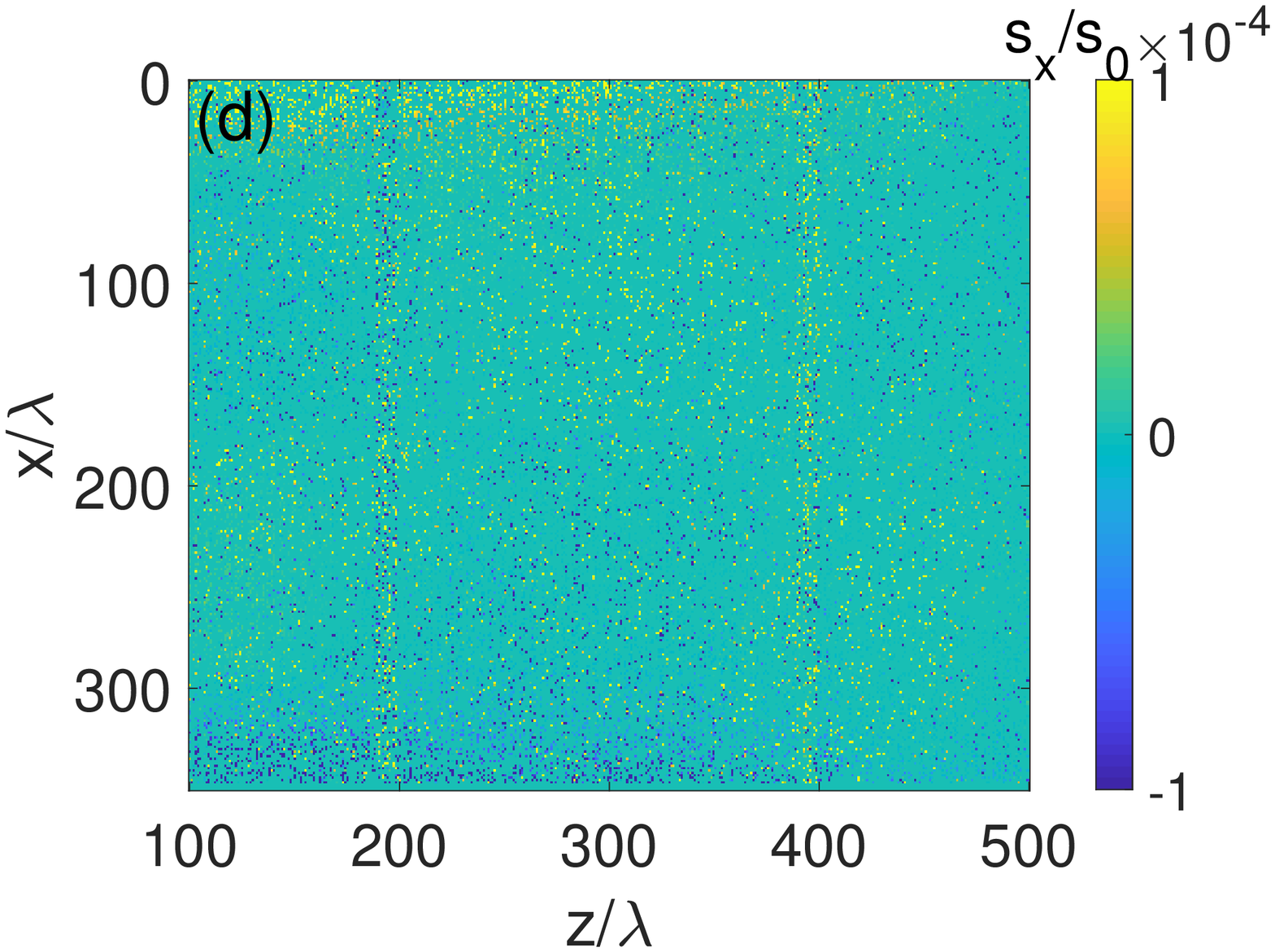}
	\caption{(a) The $s_z$ distribution when the magnetic fields are set to be zero. (b) The $s_x$ distribution when the magnetic fields are set to be zero. (c) The $s_z$ distribution when the electric fields are set to be zero. (d) The $s_x$ distribution when the electric fields are set to be zero.\label{fig:noeb}}
\end{figure}

Now we study the wakefield reconstruction from the transmitted spin distribution, i.e. getting the field intensity quantitatively from the spin distribution. For this, we firstly compared the contribution of the electric and magnetic fields on the spin precession. According to the data from the PIC simulation, one can roughly estimate their relative contribution as:
\begin{equation}
	\begin{split}
		r_{em}&=\frac{\left|\bm\Omega_B\right|}{\left|\bm\Omega_E\right|} \\
		&=\frac{\left|\bm B\left(\frac{1}{\gamma}+ a_e\right)-\frac{\gamma}{\gamma+1}\frac{\bm v}{c^2}\left(\bm v\cdot\bm B\right)\right|}{\left|\frac{\bm v}{c^2}\times\bm E\left(\frac{1}{\gamma+1}-a_e\right)\right|} \\
		&\lesssim 0.1
	\end{split}
	\label{eq:EBcompare}
\end{equation}
As one can see the spin precession is mainly induced by the electric fields. However since the directions of electric and magnetic fields are both changing when the probe beam travels through the wakefield, a careful judgement is still needed. For this, we use our spin precession program and the data from PIC simulations to calculate the spin precession in two different cases. In the first case we set the magnetic fields felt by the electrons to be zero. The spin distribution of the transmitted probe is shown in Fig.~\ref{fig:noeb}(a,b). One can still see the wake-like image. However if we set the electric fields to be zero and use magnetic fields only, the wake-like image disappears as shown in Fig.~\ref{fig:noeb}(c,d). Therefore, we can make sure that the electric fields dominate the spin precession and it is reasonable to neglect the contribution from the magnetic fields in the following fields reconstruction. Certainly it also means that we can only get the electric fields information in the current probe case.

Since the probe is relativistic and the wake size is limited, the time for transmission through the wake field is too short to gain enough transverse kicking\cite{densitydiag}. We then assume that the velocity of the probe beam does not change by the wakefields while it travels through the wake. After substituting the velocity of the probe beam to the T-BMT equation and normalize the equation to remove the constant coefficients, one can obtain the total angular velocity of spin precession and the spin evolution as:
\begin{gather}
	\bm\Omega=\bm\Omega_T+\bm\Omega_a=\frac{ev}{mc^2}\left[-E_z\left(a_e+\frac{1}{\gamma+1}\right),0,E_x\left(a_e+\frac{1}{\gamma+1}\right)\right] \\
	\frac{\diff\bm s}{\diff t}=\left[\Omega_ys_z-\Omega_zs_y,\Omega_zs_x-\Omega_xs_z,\Omega_xs_y-\Omega_ys_x\right]
\end{gather}

Moreover, since the spin evolution satisfies $s_y\approx 1,s_{x,z}\approx 0$ during the whole precession, one can simplify the spin evolution as:
\begin{gather}
	\frac{\diff s_x}{\diff t}\approx-\Omega_z \\
	\frac{\diff s_z}{\diff t}\approx\Omega_x	\label{eq:diff}
\end{gather}
By integrating Eq.~(\ref{eq:diff}) one can obtain:
\begin{equation}
	\Delta s_z=\int\Omega_x\diff t= -\frac{ev}{mc^2}\left(a_e+\frac{1}{\gamma+1}\right)\int E_z\diff t
	\label{eq:moveint}
\end{equation}

However, in Eq.~(\ref{eq:moveint}) one should integrate a moving electric field since the wake is moving at the speed of light when the probe beam travels through it. To discuss the difference between a moving and a static wakefield, we take the linear wakefield as an example. The linear wake usually has the following profile:
\begin{equation}
	E_z\sim e^{-\frac{x^2+ y^2}{2\sigma_p^2}}\cos k_p\left(z-v_p t+\phi_0\right)
\end{equation}
where $v_p$ is the phase velocity of the wakefield and $\phi_0$ is the initial phase. For the field detection, $\phi_0$ actually is different for each probe electrons since they touch the field at different time. However, in our case, the thickness of the probe beam ($d=2\mu m$ ) is thin enough compared with the wake period length ($\lambda_w\approx 100\mu m$), so $\phi_0$ can be regarded as 0 for all the probe electrons. Since the probe beam and wake are both moving with a speed close to the light speed $c$, $v$ and $v_p$ can be both regarded as 1 after normalization. In addition, the probe beam travels long enough, so we can take the upper and lower limit of the integration to be infinity. In this way one can easily get the effective part of the integration in a static field case($v_p=0$) by using $\diff t=\diff y/v\approx\diff y/c$ as:
\begin{equation}
	I_s=\int_{-\infty}^{+\infty}e^{-\frac{y^2}{2\sigma_p^2}}\cos k_pz\diff y= \sqrt{2\pi}\sigma_p \cos k_pz
\end{equation}

For the moving wake($v_p=1$), one can get the effective part of the integration as:
\begin{equation}
			\begin{split}
				I_m&=\int_{-\infty}^{+\infty}e^{-\frac{y^2}{2\sigma_p^2}}\cos k_p(z-y)\diff y \\
				&=\int_{-\infty}^{+\infty}e^{-\frac{y^2}{2\sigma_p^2}}\left(\cos k_pz\cos k_py+\sin k_pz\sin k_py\right)\diff y \\
				&=\int_{-\infty}^{+\infty}e^{-\frac{y^2}{2\sigma_p^2}}\cos k_pz\cos k_py\diff y
			\end{split}
		\end{equation}
This can be further simplified as:
\begin{equation}
		\begin{split}
			I_m&=\text{Re}\left(\cos k_pz\int_{-\infty}^{+\infty}e^{-\frac{y^2}{2\sigma_p^2}+ ik_py}\diff y\right) \\
			&=\text{Re}\left[\cos k_pz\int_{-\infty}^{+\infty}e^{-\frac{(y-ik_p\sigma_p^2)^2}{2\sigma_p^2}-\frac{k_p^2\sigma_p^2}{2}}\diff y\right] \\
			&=\sqrt{2\pi}\sigma_p\cos k_pz e^{-\frac{k_p^2\sigma_p^2}{2}}
		\end{split}
\end{equation}

By comparing this with the static field result, one can know that they just differ by a factor of $\exp{(-k_p^2\sigma_p^2/2)}$:
\begin{equation}
	I_m= e^{-\frac{k_p^2\sigma_p^2}{2}}I_s
\end{equation}

\begin{figure}[ht]
	\centering
	\includegraphics[scale=0.21]{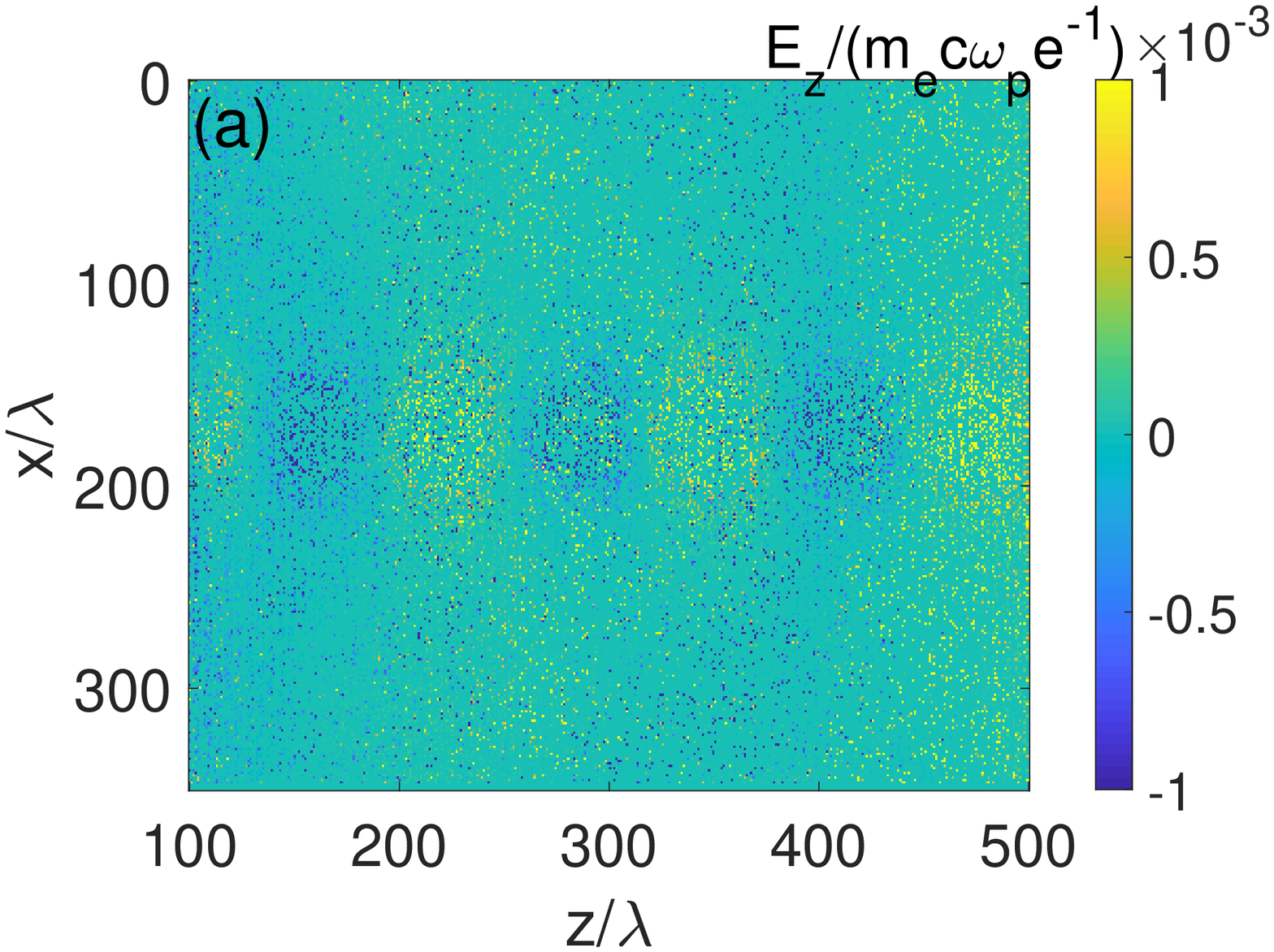}
	\includegraphics[scale=0.21]{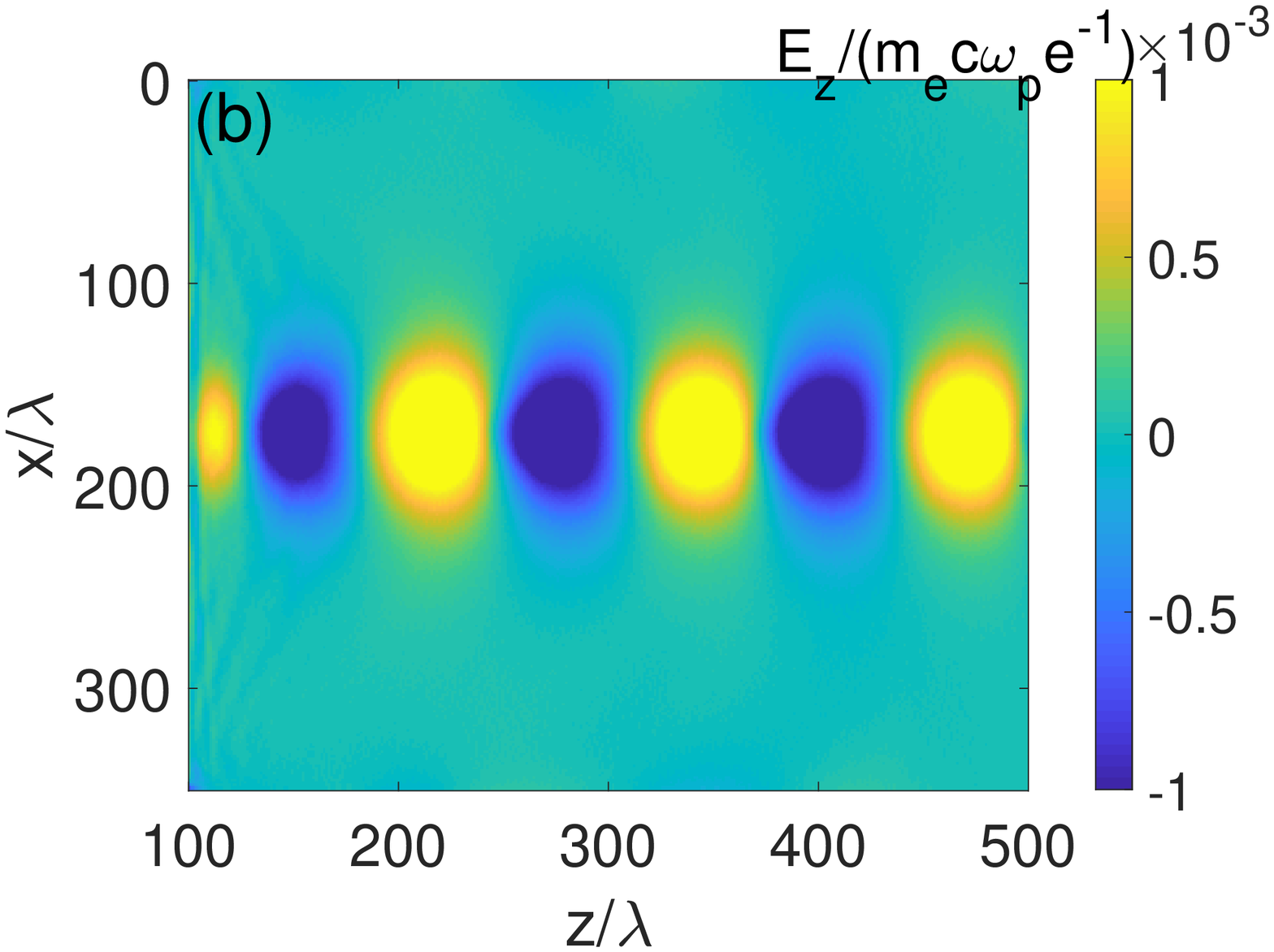}
	\includegraphics[scale=0.21]{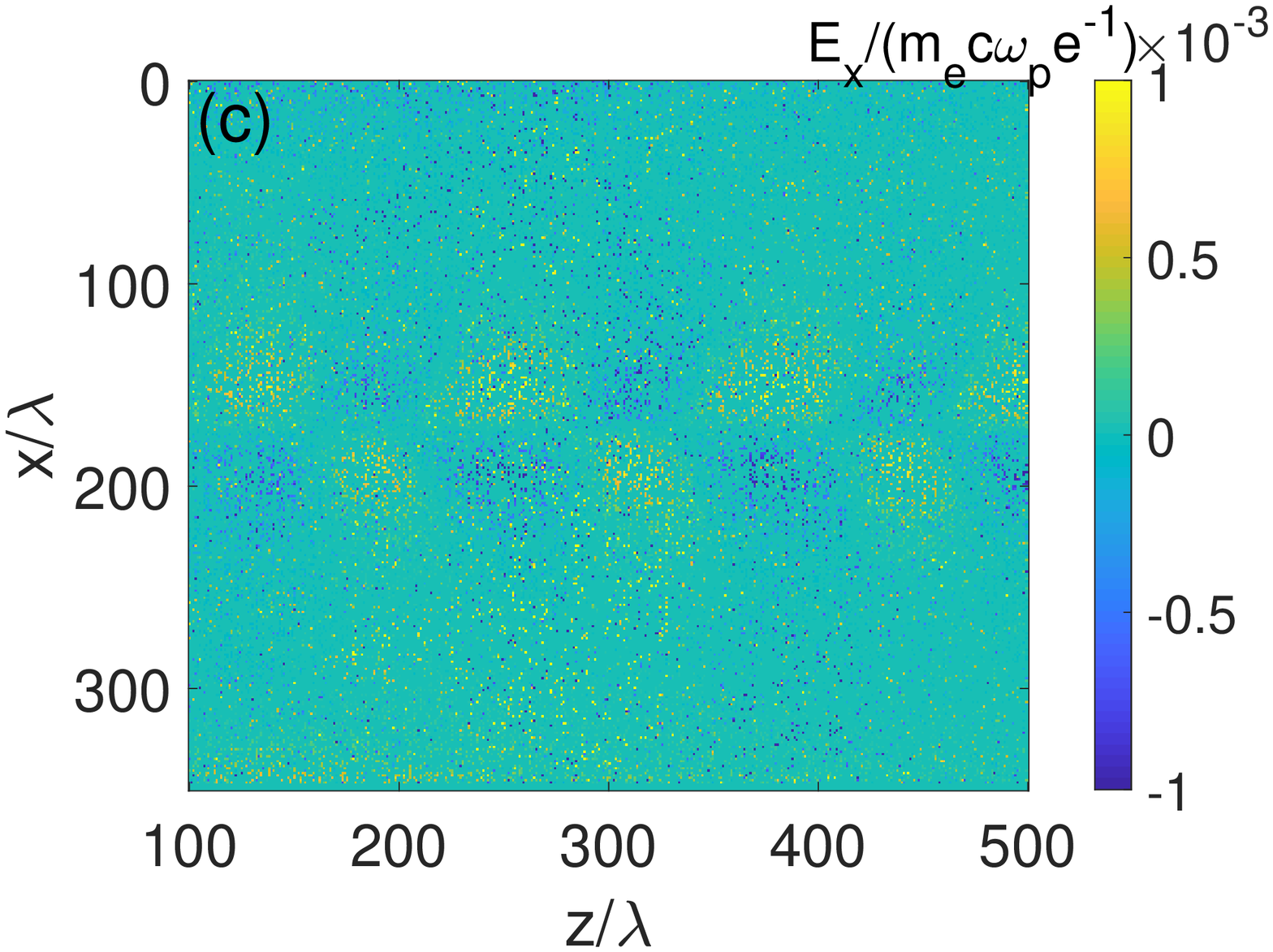}
	\includegraphics[scale=0.21]{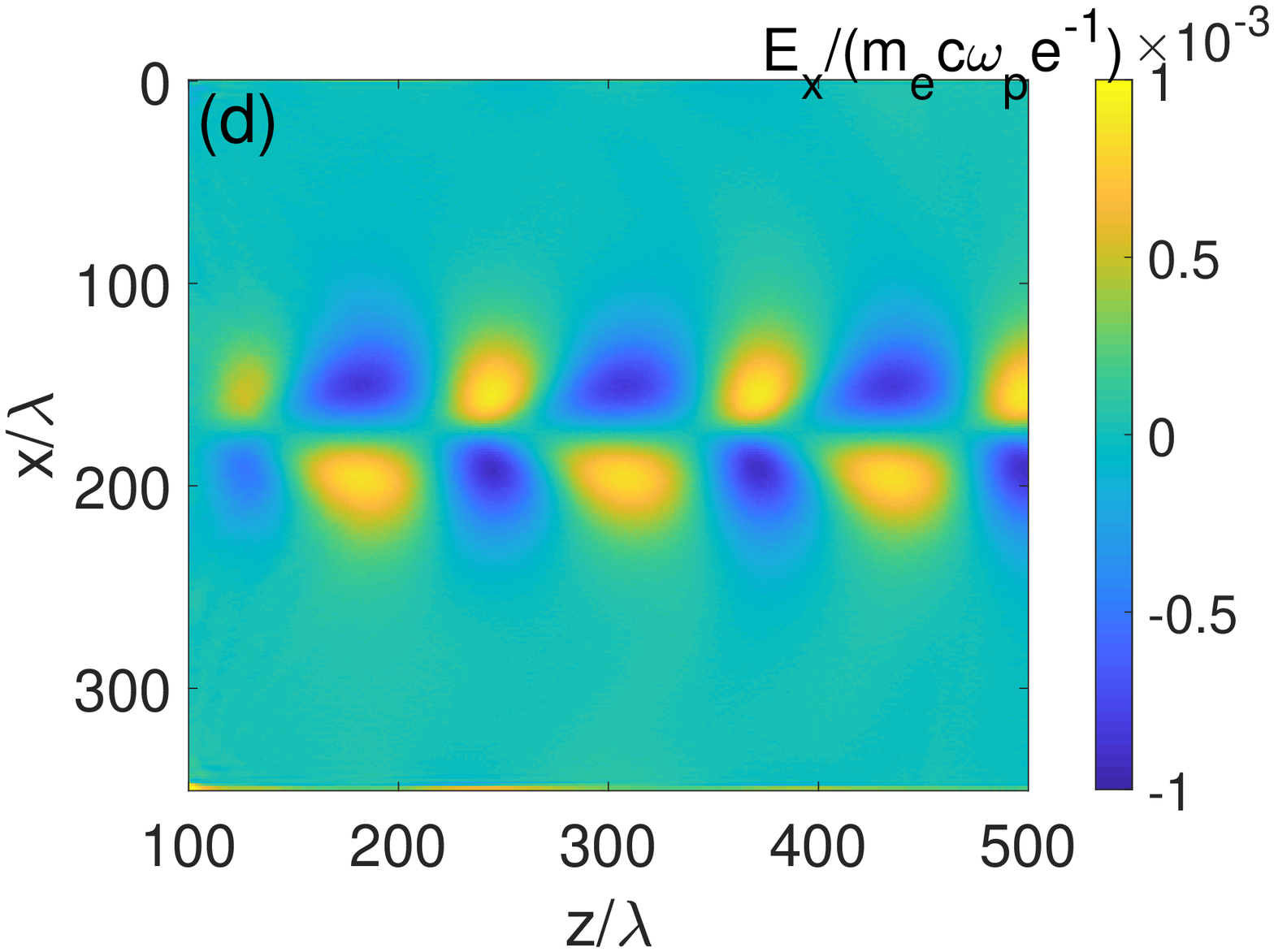}
	\caption{(a) The longitudinal wakefield $E_z$ reconstructed from spin evolution. (b) The $E_z$ field calculated in PIC simulation. (c) The transverse wakefield $E_x$ reconstructed from spin evolution. (d) The $E_x$ field calcualted in PIC simulation.\label{fig:compare}}
\end{figure}

 Therefore, one can obtain the electric fields in the wakefield from the variation of spins as:
\begin{gather}
	\bar E_z=-e^{\frac{k_p^2\sigma_p^2}{2}}\left(a_e+\frac{1}{\gamma+ 1}\right)^{-1}\frac{mc^2\Delta s_z}{ev\Delta t} \\
	\bar E_x=-e^{\frac{k_p^2\sigma_p^2}{2}}\left(a_e+\frac{1}{\gamma+ 1}\right)^{-1}\frac{mc^2\Delta s_x}{ev\Delta t}
\end{gather}
where $\Delta t$ is the time for the probe beam to transmit through the wake. The comparison between the electric fields reconstructed from the spin evolution and the PIC simulation are shown in Fig.~\ref{fig:compare}. As we can see the reconstruction fields reproduce the main characters of the original fields.

It deserves to point out that although in our current study we have used a layered probe beam with thickness of $d=2\,\mu$m, which is usually difficult to get. For a thicker probe beam, it can be regarded as a series of thin layers. Then the final image and reconstructed field are also superpositions of the results of these thin probe beams. Considering the influence of $\phi_0$, to get distinguishable results, one can see that the thickness of the beam should satisfy $d<\lambda_w$. In further, in our study to speed up the simulation, we have just simulated about $20\%$ of the macro-particles from the PIC simulation. This can be viewed as a sample from the probe beam along transverse section. The more probe particles used, the smoother the reconstructed fields will be.

\subsection{Reconstruction of field structures in the transverse directions}

\begin{figure}[ht]
	\centering
	\includegraphics[scale=1.6]{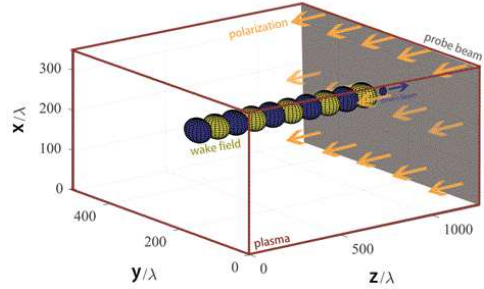}
	\caption{\label{fig:schemeparallel}The spin polarization diagnostic scheme in antiparallel direction.}
\end{figure}

Besides the reconstruction of the field structures in the longitudinal direction, the transverse structures can also be detected by using a counter propagating probe beam. As shown in Fig.~\ref{fig:schemeparallel}, the driver beam and background plasma are the same as the previous case, but the probe beam is incident to the plasma in the opposite direction. The energy of the electrons of the probe beam is still 200\,MeV and polarized along the moving direction $-z$.

\begin{figure}[ht]
			\centering
			\includegraphics[scale=0.21]{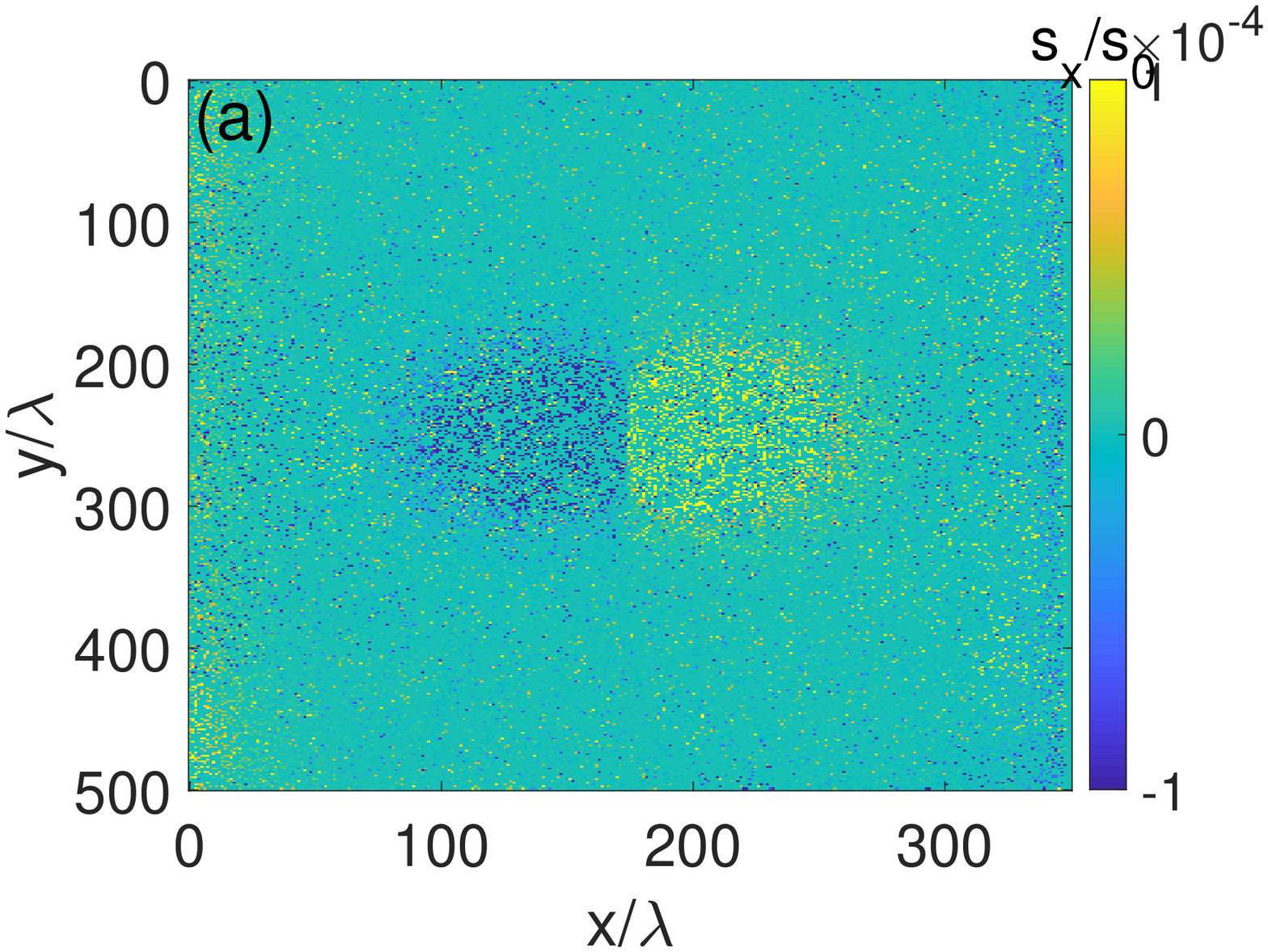}
			\includegraphics[scale=0.21]{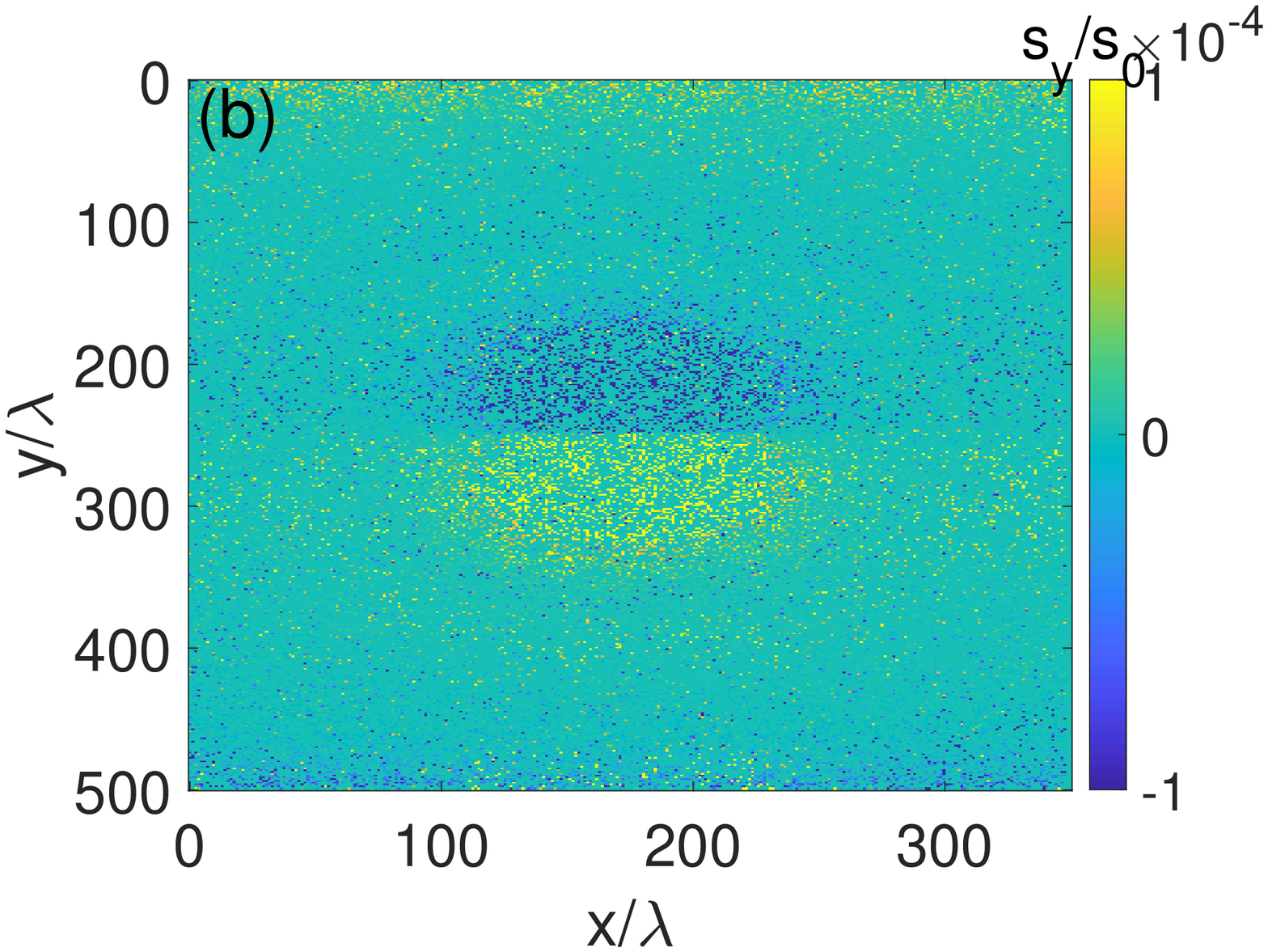}
			\caption{\label{fig:distparallel}(a)The distribution of $s_x$ in the $x$-$y$ plane after the probe beam transmits through the wakefields. (b)The distribution of $s_y$ in the $x$-$y$ plane after the probe beam transmits through the wakefields.}
		\end{figure}

\begin{figure}[ht]
			\centering
			\includegraphics[scale=0.21]{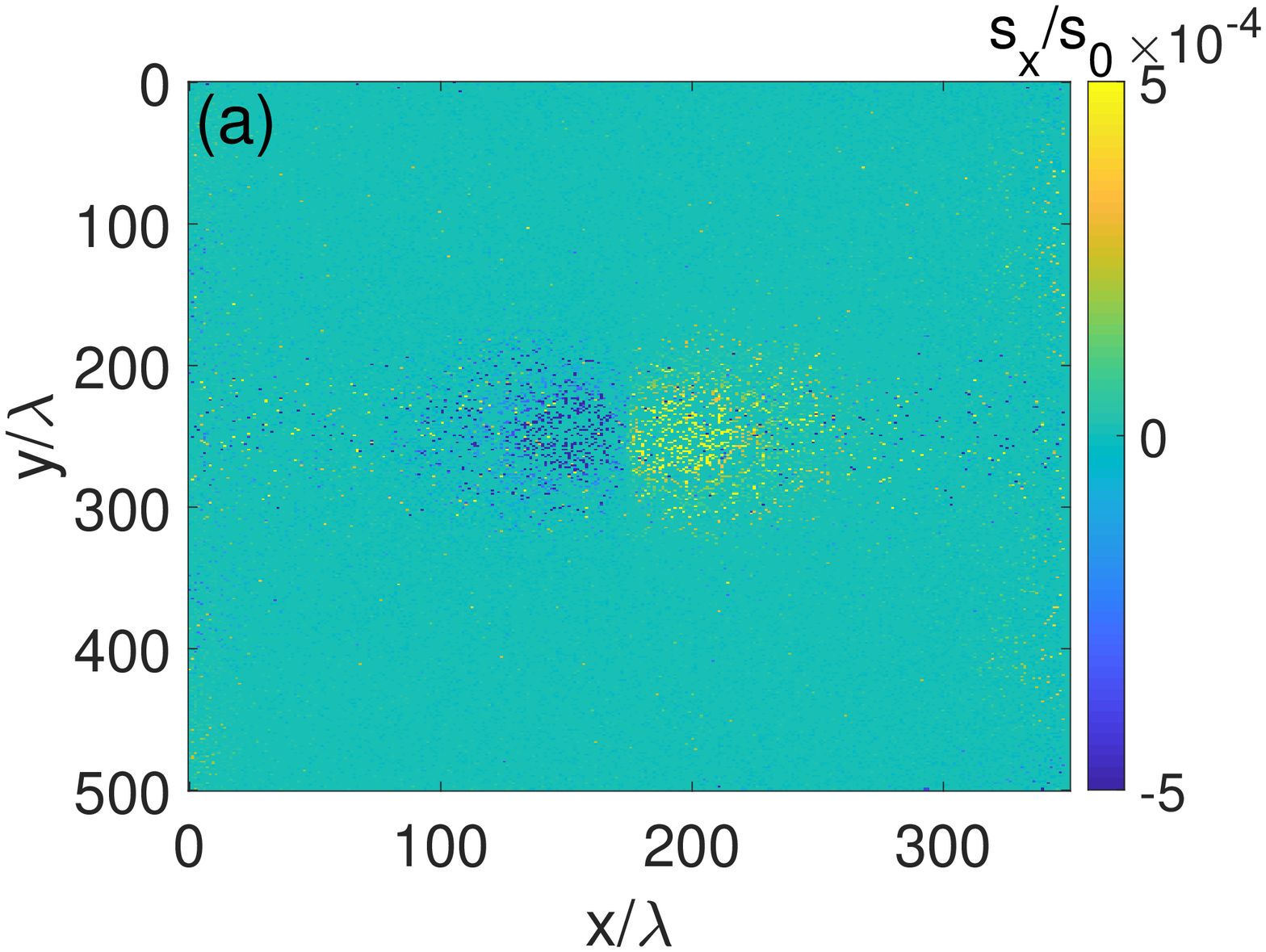}
			\includegraphics[scale=0.21]{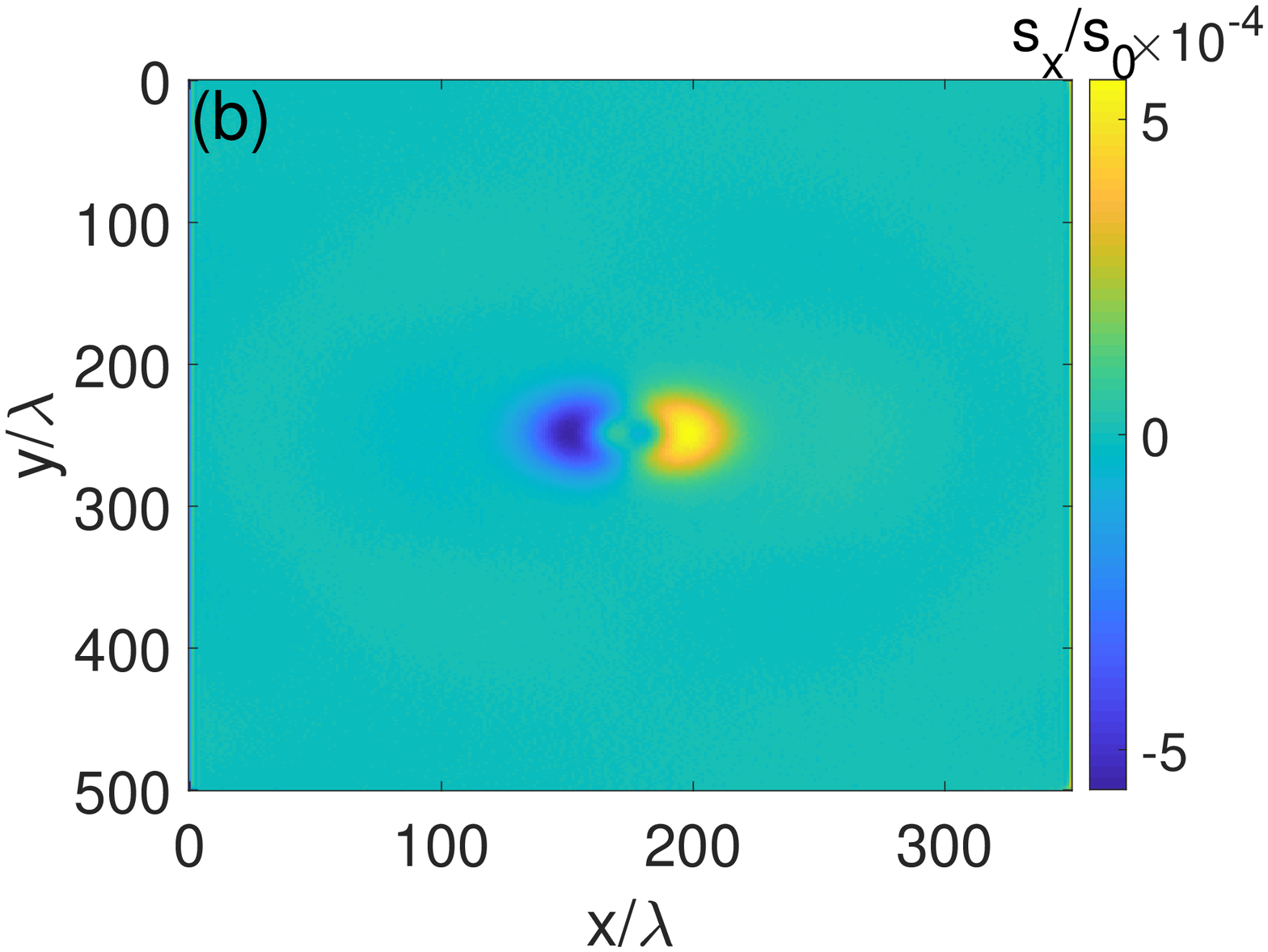}
			\includegraphics[scale=0.21]{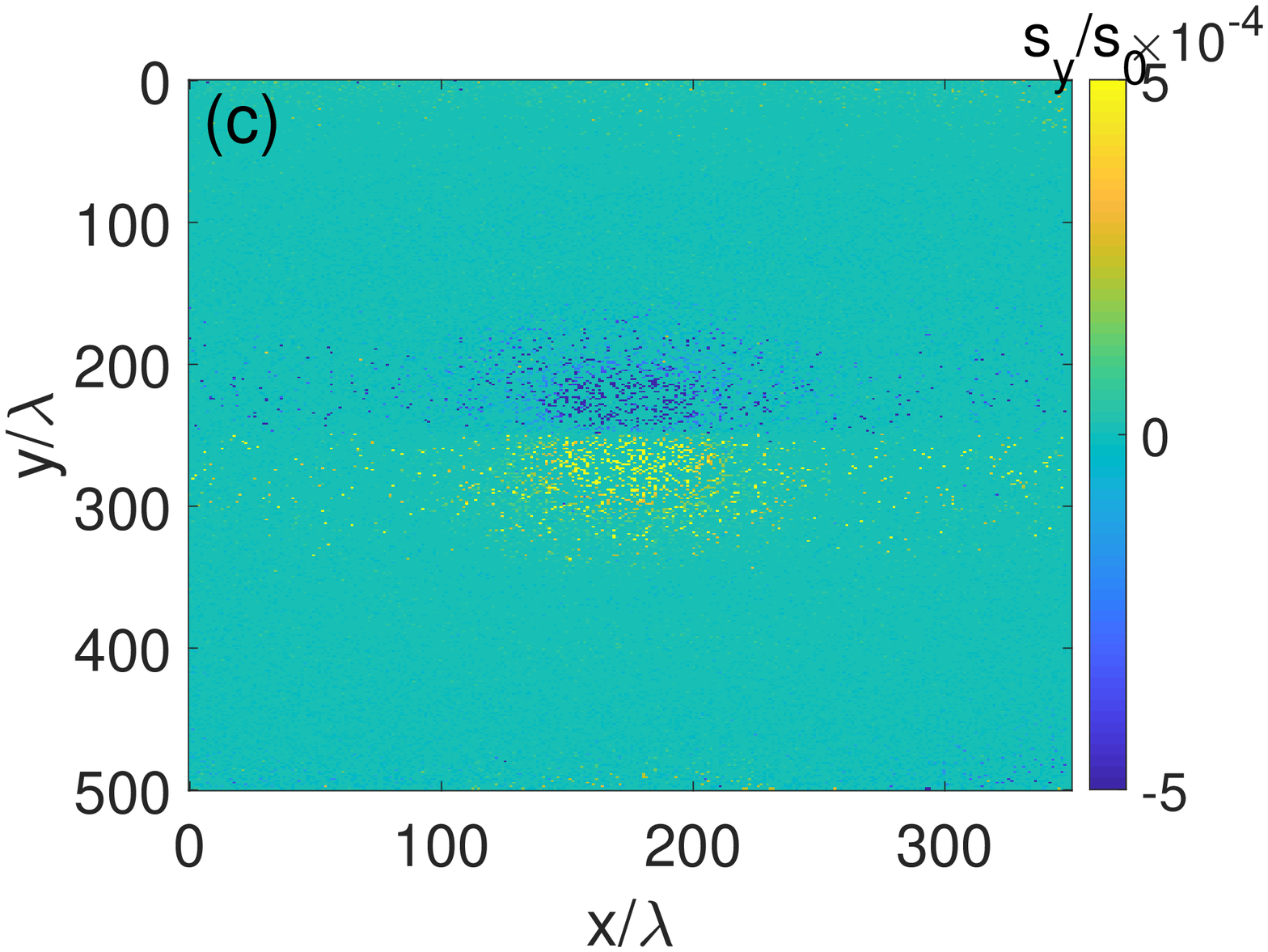}
			\includegraphics[scale=0.21]{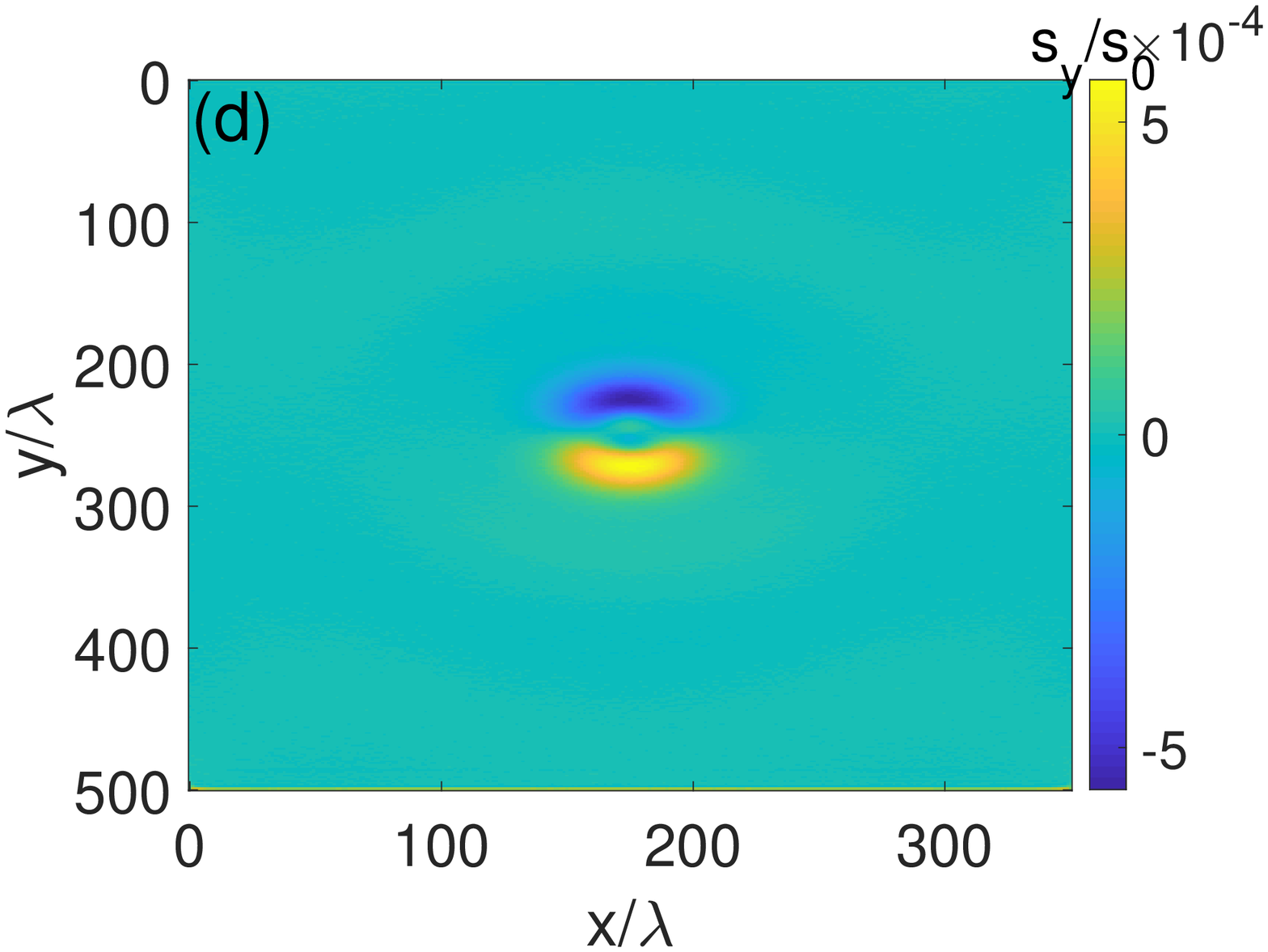}
			\caption{\label{fig:reverseparallel}(a)The transverse wakefield $E_x$ reconstructed from spin evolution. (b)The $E_x$ field calculated in PIC simulation (c)The transverse wakefield $E_y$ reconstructed from spin evolution. (d) The $E_y$ field calculated in PIC simulation.}
		\end{figure}

As before the spin distribution after the probe beam transmitting through the wakefields has been recorded as shown in Fig.~\ref{fig:distparallel}. Through similar assumption and analysis we can get the relationship between the electric field in the wakefields and the spin evolution, as described by Eq.~(\ref{eq:parallelreverse}) .

\begin{gather}
	\bar E_y=\left(a_e+\frac{1}{\gamma+ 1}\right)^{-1}\frac{mc^2\Delta s_y}{ev\Delta t} \notag\\
	\bar E_x=\left(a_e+\frac{1}{\gamma+ 1}\right)^{-1}\frac{mc^2\Delta s_x}{ev\Delta t} \label{eq:parallelreverse}
\end{gather}
The comparison between the reconstructed electric fields from spin evolution and the PIC simulation fields are shown in Fig.~\ref{fig:reverseparallel}. As one can see the main characters of the transverse fields have been reconstructured.

\section{effects of the probe beam's energy}
Although in principle the plasma fields can be reconstructed from spin evolution, the resolution of the spin evolution i.e. the sensitivity of the spin detector, is obviously a critical factor to determine the usefulness of such scheme. Nowadays, there are several ways to measure the spin polarization. One is so called Mott polarimetry and in this way one obtains the polarization of an electron beam through transverse scattering by targets made of high Z atoms\cite{polarimetry}. One can also use an optical polarimetry scheme where polarized electrons excite atomic target to high order states and then radiate circularly polarized fluorescence\cite{polarimetry}. By using these methods, the variation of the beam spin polarization \cite{precession1,precession2,precession3,precession4} in the level of $\lesssim10^{-2}$ can be measured. However in our scheme the spin variation is in the $10^{-4}$ level, a higher resolution spin detector should be used.

\begin{figure}
	\centering
	\includegraphics[scale=0.21]{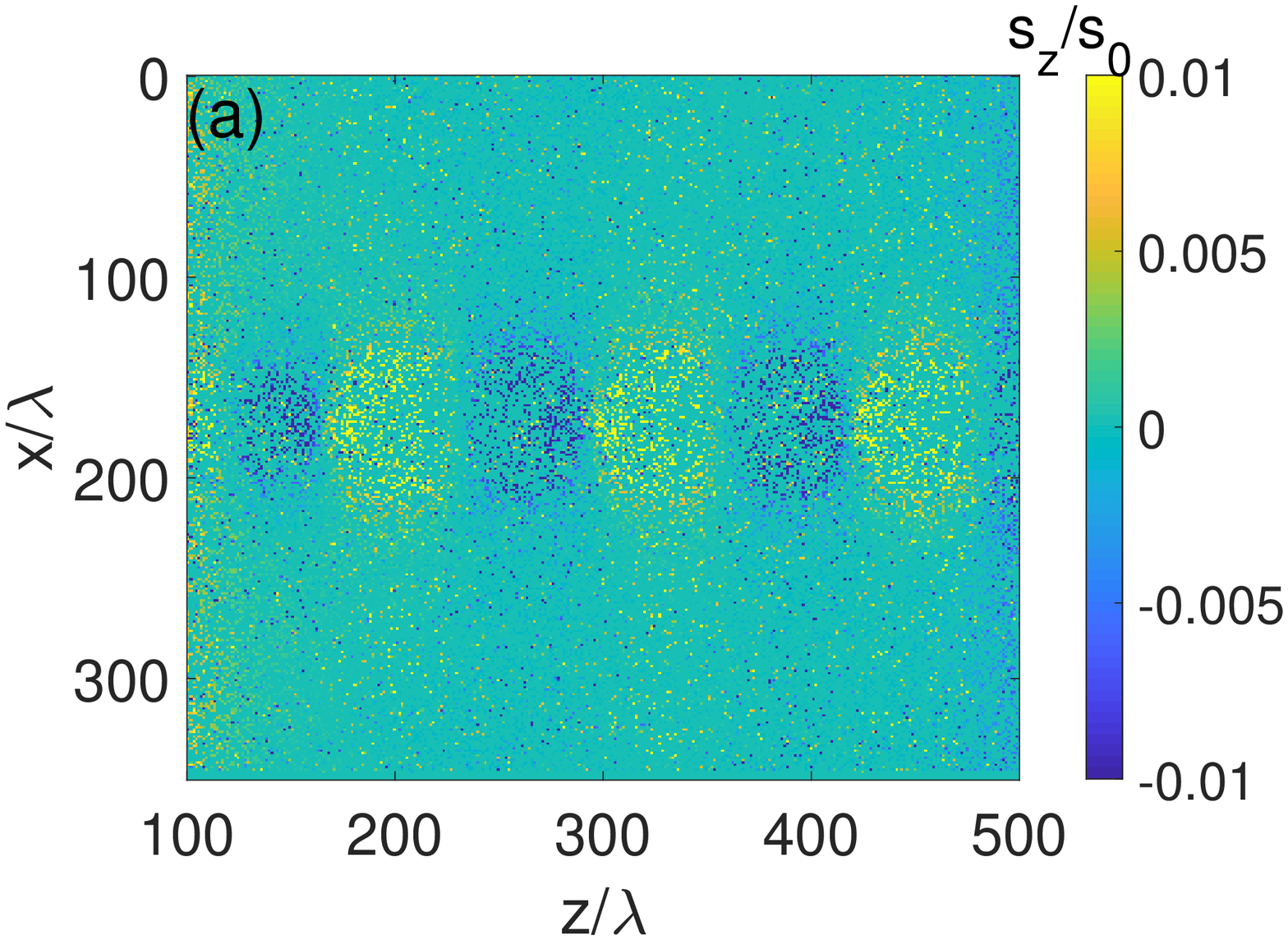}
	\includegraphics[scale=0.21]{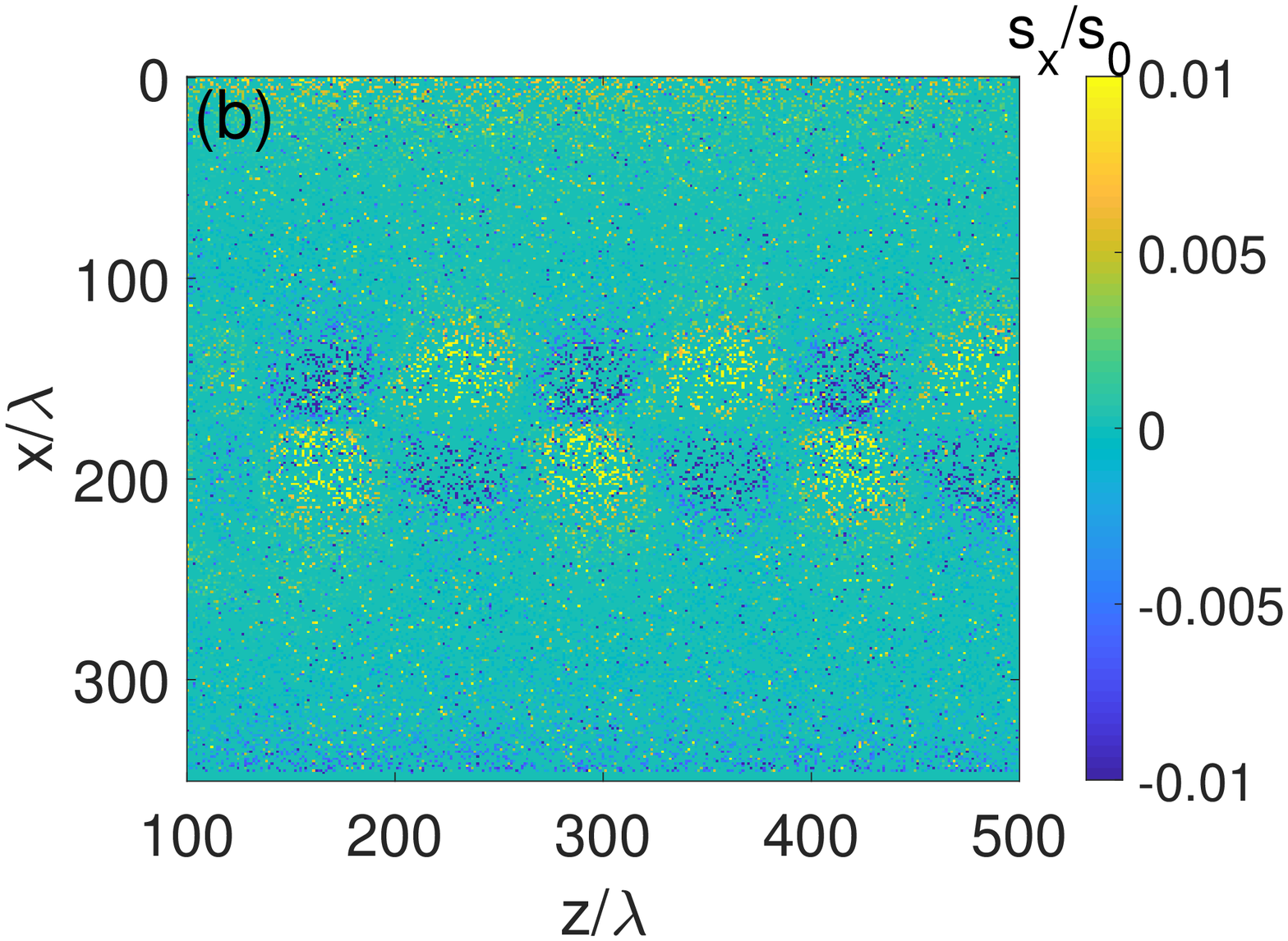}
	\includegraphics[scale=0.21]{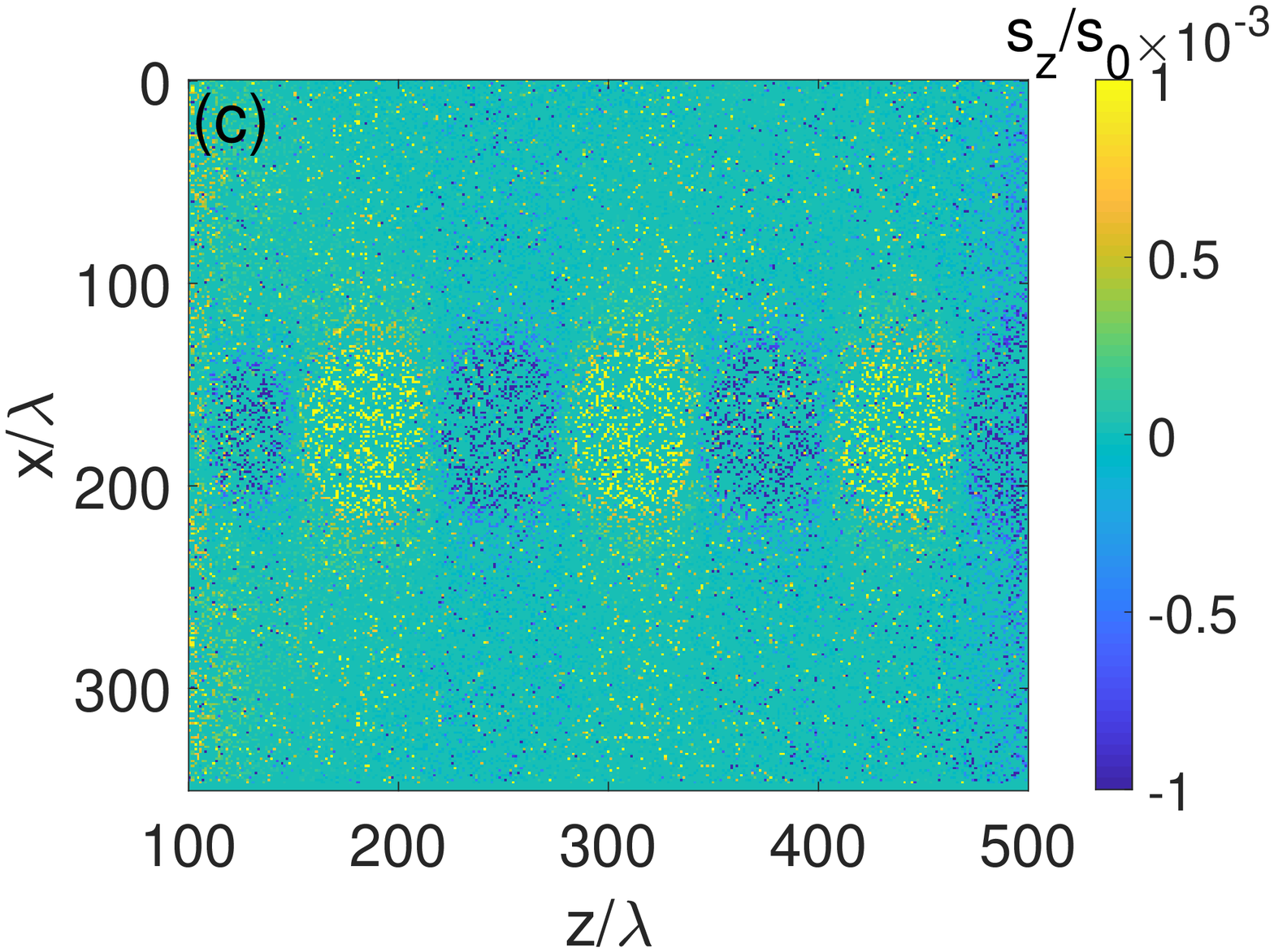}
	\includegraphics[scale=0.21]{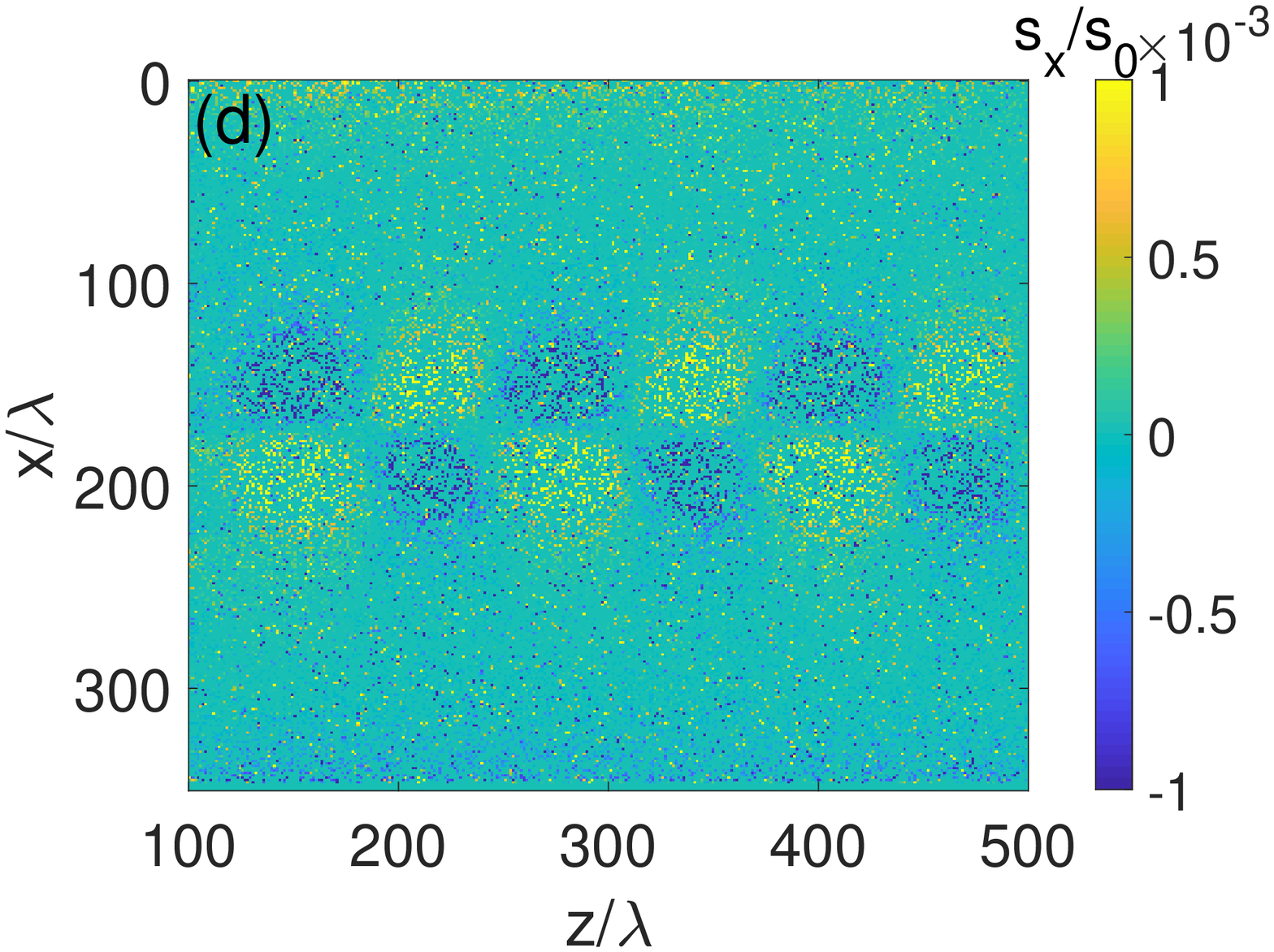}
	\caption{The distribution of $s_z$ (a) and $s_x$ (b) when a probe beam with electron energy of 2.6\,MeV is used.  The distribution of $s_z$ (c) and $s_x$ (d) when a probe beam with electron energy of 26\,MeV is used.  \label{fig:lowE}}
\end{figure}

To overcome this, we have studied the effects of the probe beam energy and found that a probe beam with lower energy can be used to reduce the requirement on the sensitivity of the spin detector. This is because the precession frequency $\bm\Omega_T$ increases as the energy of the probe beam decreases. The variation of the spin is much larger for a low energy probe beam. We checked this by using probe beams with electron energy of 2.6\,MeV and 26\,MeV (the relativistic momentums after normalization are 5 and 50 respectively). The spin distributions after probe beam transmission are shown in Fig.~\ref{fig:lowE}. As one can see for the 2.6\,MeV case, the spin variation is in the level of $1\%$ which is detectable by today's spin detector. However, we should say for intense fields case, the movement of the probe in the fields maybe too large, the low energy electron beam will be deviated during the transmission. Only high energy probe beam can carry the spin information along the detected fields, then due to the smaller variation of the spin, a spin detector with higher sensitivity is still needed.

\section{Discussion and summary}
Besides the high requirement of the detector's sensitivity on spin evolution, the spatial resolution of the spin distribution is also critical. Currently we have reconstructed the fields directly from the image of the spatial distribution of the transmitted electron beam. In practice, this image may be amplified by normal electron beam optical devices. The enlarged electron beam can then be detected by arrays of spin detectors. The spatial spin revolution is determined by the beam optics and the spin evolution sensitivity is determined by a single spin detector. Certainly the spin evolution along the transported beam line should be took into account during the field reconstruction. Detailed beam optics design is outside the scope of our current study.

At last we compare such spin polarized electron beam probe (SPEBP) scheme with the normal charged particle beam probe (CPBP) scheme. As one may think, for many field structures, by using only density modulation of a transmitted charged beam may not be enough to reconstruct the fields. In the normally charged particle detection scheme, the electric field in the wakefield modulates the momentum when the probe beam transmits through it and the momentum modulation changes into density modulation after the probe beam drifts a distance. However, along the probe beam's propagation direction, since the initial momentum of the electron is far larger than the modulation, one can hardly obtain such density modulation along this direction, so the electric field in this direction cannot be detected. However, as we have derived before, the electric field in each direction just corresponds to the spin distribution in each direction. Therefore we are able to get all the electric fields through spin distribution.

In a summary, we have proposed a new scheme to probe a plasma wakefield by a spin polarized electron beam and demonstrated it through numerical simulations. For some special cases, this method maybe used to differentiate the electric and magnetic fields. The scheme may also be extended to a Spin Polarized Electron Beam Computed Tomography (SPEB-CT) by using probe beams with different spin polarizations along different injection directions to mapping fields with even complex structures where the normal CPBP is not possible.

\begin{acknowledgments}
This work was supported by National Natural Science Foundation of China (Grant Nos. 11774227 and 11721091), Science Challenge Project (Grant No. TZ2018005) and National Key Research and Development Program of China (Grant No. 2018YFA0404802). The authors would like to acknowledge the OSIRIS-Consortium, consisting of UCLA and IST (Lisbon, Portugal) for the use of OSIRIS and the visXD framework.
\end{acknowledgments}


\bibliography{polarizationdiag_new}

\end{document}